\documentclass{autart}
\usepackage[dvipsnames]{xcolor}

\usepackage{natbib}        
\usepackage{amsmath}

\usepackage{amsfonts}
\usepackage{graphicx}
\usepackage{nicefrac}
\usepackage{mathrsfs}
\usepackage{upgreek}
\usepackage{todonotes}
\usepackage{subfigure}
\usepackage{epstopdf}

\definecolor{LB}{RGB}{65,105,225}

\newcommand\xqed[1]{%
  \leavevmode\unskip\penalty9999 \hbox{}\nobreak\hfill
  \quad\hbox{#1}}

\newcommand{\R}{\mathbb{R}}

\newcommand{\B}{\mathcal{B}}

\newcommand{\E}{\mathcal{E}}
\newcommand{\G}{\mathcal{G}}

\newcommand{\BS}{\mathbb{S}}
\newcommand\demo{\xqed{$\square$}}

\newtheorem{theorem}{Theorem}[section]

\newtheorem{remark}{Remark}[section]
\newtheorem{proposition}{Proposition}[section]

\newtheorem{example}{Example}[section]
\newtheorem{definition}{Definition}[section]
\newtheorem{problem}{Problem}[section]

\begin{document}
\begin{frontmatter}

\title{A Characterization of All Passivizing Input-Output Transformations of a Passive-Short System}


\author[KTH]{Miel Sharf},~ 
\author[Technion]{Daniel Zelazo} 

\address[KTH]{KTH Royal Institute of Technology, Stockholm, Sweden (e-mail: sharf@kth.se).}
\address[Technion]{Technion - Israel Institute of Technology, 
   Haifa, Israel (e-mail: dzelazo@technion.ac.il).}

\begin{abstract}                
Passivity theory is one of the cornerstones of control theory, as it allows one to prove stability of a large-scale system while treating each component separately. In practice, many systems are not passive, and must be passivized in order to be included in the framework of passivity theory. Input-output transformations are the most general tool for passivizing systems, generalizing output-feedback and input-feedthrough. In this paper, we classify all possible input-output transformations that map a system with given shortage of passivity to a system with prescribed excess of passivity. We do so by using the connection between passivity theory and cones for SISO systems, and using the S-lemma for MIMO systems. We also present several possible applications of our results, \textcolor{black}{including simultaneous passivation of multiple systems or with respect to multiple equilibria, as well as optimization problems such as $\mathcal{L}_2$-gain minimization. We also exhibit our results in a case study about synchronization in a network of non-passive faulty agents.}
\end{abstract}


\end{frontmatter}



\section{Introduction}
Over the last few decades, many engineering systems have become much more complex, as networked systems and large-scale systems turned common, and ``system-of-systems" evolved into a leading design methodology. To address the ever-growing complexity of systems, many researchers suggested various component-level tools that guarantee system-level properties, e.g. input-output stability. One important example of such a notion is passivity, which can be informally stated as ``energy-based control" [\cite{VanDerSchaft1999}]. It has been used to solve different problems in control theory in many areas, including networked systems [\cite{Arcak2007,Bai2011}], cyber-physical systems [\cite{Antsaklis2013}], robotics [\cite{Hatanaka2015}] and power systems [\cite{DePersis2018}].

In practice, however, many systems are not passive. Examples include systems with input/output delays (such as chemical processes), human operators, generators, and power networks, among others  [\cite{Trip2018,Xia2014,Harvey2016,Atman2018}]. The lack of passivity is often quantified using passivity indices. In order to use passivity-based design techniques, one needs to passivize the system under consideration. The most common methods for passivation include gains, output-feedback, input-feedthrough, or a combination thereof [\cite{Byrnes1991,Zhu2014,Jain2018,Sharf2019c}]. 

More generally, a passivation method relying on an input-output (I/O) transformation was suggested in \cite{Xia2018}. An I/O transformation is a concise formulation aggregating output-feedback, input-feedthrough, and gains. Namely, \cite{Xia2018} \textcolor{black}{generalized the well known Cayley Transform [\cite{VanDerSchaft1999}] to show} that any system with finite $\mathcal{L}_2$-gain can be passivized using an I/O transformation, found algebraically. More recently, \cite{Sharf2019a} used a geometric approach to prescribe a passivizing I/O transformation for SISO systems. More precisely, one constructs an I/O transformation, mapping a system with known passivity indices to a system with prescribed passivity indices. This was achieved using a connection between passivity and cones through the notion of projective quadratic inequalities (PQIs), \textcolor{black}{which can be seen as a specific case of sector bounds [\cite{Safonov1982}]}.

In this paper, we use the geometric approach of \cite{Sharf2019a} to give a full description of all passivizing I/O transformations of a given SISO system. More precisely, we give a concise description of all I/O transformations that map a system with known passivity indices to a system with prescribed excess of passivity. This is done by understanding the action of the group of (invertible) I/O transformations on the collection of cones in the plane, \textcolor{black}{which is a byproduct of the geometric approach of \cite{Sharf2019a}, allowing us to use standard group theory methods.} We show that any transformation mapping a system with known passivity indices to a system with prescribed excess of passivity can be written (up to a scalar) as a product of three matrices - one depending on the original passivity indices, one depending on the desired excess of passivity, and a matrix whose entries are all non-negative. We then use similar mechanisms to give an analogous result for MIMO systems, where the non-negative matrix is replaced by a matrix satisfying a certain generalized algebraic Riccati inequality. 

Our results can be seen as an analogue of the Youla parameterization [\cite{Kucera2011}], dealing with passivizing I/O transformations instead of stabilizing controllers. We also consider multiple application domains of our results. First, we consider transformations simultaneously passivizing multiple systems, and explore applications in fault mitigation and plug-and-play control. Second, we consider the problem of passivizing a plant with respect to multiple equilibria, which relates to equilibrium-independent passivity [\cite{Hines2011}]. \textcolor{black}{Lastly, we consider the problem of finding a passivizing transformation $T$ which optimizes a certain cost function, e.g. the $\mathcal{L}_2$-gain of the transformed system, or the distance in the operator norm between the original system and the transformed system. We demonstrate our findings by a case study about synchronization in networks with faulty and non-passive agents.}

The rest of the paper is organized as follows. Section \ref{Sec.background} presents the geometric approach of \cite{Sharf2019a} and formulates the problem. Section \ref{sec.SISO} characterizes all passivizing transformations of a given passive-short SISO system, and Section \ref{sec.MIMO} generalizes the characterization to MIMO systems. Sections \ref{sec.AppExam} \textcolor{black}{and \ref{sec.Synch}} provide possible applications of the achieved results.

\paragraph*{Notation:}
We denote the group of all invertible matrices $T \in \R^{d\times d}$ as $GL_d(\R)$. Given a linear transformation $S:\R^d \to \R^d$ and a basis $\B$ for $\R^d$, we denote the representing matrix of $S$ in the basis $\B$ as $[S]_\B$. Furthermore, given two bases $\B_1,\B_2$ of $\R^d$, we denote the change-of-base matrix from $\B_1$ to $\B_2$ by $I_{\B_1\to\B_2}\in GL_d(\R)$. We note that $I_{\B_1\to\B_2}^{-1} = I_{\B_2\to\B_1}$. Moreover, for any linear transformation $S:\R^d\to\R^d$, we have that $[S]_{\B_2} = I_{\B_1\to\B_2}[S]_{\B_1}I_{\B_2\to\B_1}$. We also denote the Kronecker product by $\otimes$, and the $d\times d$ identity matrix as ${\rm Id}_d$. Lastly, we denote the unit circle inside $\R^2$ by $\BS^1$.

\section{Background and Problem Formulation} \label{Sec.background}
We consider dynamical systems given by the state-space representation $\dot{x} = f(x,u),\, y=h(x,u),$ where $u\in \R^{n_u}$ is the input, $y\in \R^{n_y}$ is the output, and $x\in \R^{n_x}$ is the state of the system. We recall the definition of passivity:

\begin{definition}
Let $\Sigma$ be a dynamical system with equal input and output dimensions. Assume that $u=0,y=0$ is an equilibrium of the system. We say the system is \emph{passive} if there exists a positive-definite $C^1$-smooth function (i.e., a storage function) $S(x)$ such that:
\begin{align}\label{eq.Passivity}
\frac{dS(x(t))}{dt} = \nabla S(x(t))\dot{x}(t) \le u(t)^\top y(t),
\end{align}
holds for any trajectory $(u(t),x(t),y(t))$ of the system.
\end{definition}

The notion of passivity stems from energy-based control, as $S(x)$ can be thought of the potential energy stored inside the system, so \eqref{eq.Passivity} implies that the change in the energy stored in the system cannot be greater than the supplied power. Passivity has been used to solve problems in various application domains, including multi-agent networks [\cite{Arcak2007,Bai2011}], cyber-physical systems [\cite{Antsaklis2013}] and robotics [\cite{Hatanaka2015}]. We can expand the notion of passivity to consider both the case of total energy dissipation, and the case of (bounded) total energy gain, by adding either a negative or a positive term to the right-hand side of \eqref{eq.Passivity}:
\begin{definition} \label{def.WPassivity}
Let $\Sigma$ be a dynamical system with equal input and output dimensions. Assume that $u=0,y=0$ is an equilibrium of the system. Let $\rho,\nu\in\mathbb{R}$.
\begin{itemize}
\item[i)] We say that the system is \emph{output $\rho$-passive} if there exists a storage function $S$ such that the inequality,
\begin{align}\label{eq.OWPassivity}
\frac{dS(x(t))}{dt} \le u(t)^\top y(t) - \rho \|y(t)\|^2,
\end{align}
holds for any trajectory $(u(t),x(t),y(t))$ of the system.
\item[ii)] We say that the system is \emph{input $\nu$-passive} if there exists a storage function $S$ such that the inequality,
\begin{align}\label{eq.IWPassivity}
\frac{dS(x(t))}{dt} \le u(t)^\top y(t) - \nu \|u(t)\|^2,
\end{align}
holds for any trajectory $(u(t),x(t),y(t))$ of the system.
\item[iii)] We say that the system is \emph{input-output ($\rho,\nu$)-passive} if $\rho\nu < 1/4$ and there's a storage function $S$ such that,
\begin{align}\label{eq.IOWPassivity}
\frac{dS(x(t))}{dt} \le u(t)^\top y(t) - \rho \|y(t)\|^2 - \nu \|u(t)\|^2,
\end{align}
holds for any trajectory $(u(t),x(t),y(t))$ of the system.
\end{itemize}
\end{definition}

\begin{rem}
The demand $\rho\nu < 1/4$ is made to assure that the right-hand side of \eqref{eq.IOWPassivity} is not always positive, nor always negative, as it would either imply that all static nonlinearities are I/O $(\rho,\nu)$-passive, or that no system is I/O $(\rho,\nu)$-passive, both of which are absurd.
\end{rem}

The case in which $\rho,\nu > 0$ is usually referred to as \emph{strict passivity} (or ``excess of passivity"), and the case in which $\rho,\nu < 0$ is usually called \emph{passive short} (or ``shortage of passivity"). The definition above allows us to consider both cases in a united framework. Passive-short and non-passive systems appear in many practical applications, see e.g. \cite{Atman2018} or \cite{Xia2014}. In order to incorporate them into passivity-based control schemes, one usually passivizes them using a transformation [\cite{Byrnes1991,Sharf2019a}]. Common transformations include output-feedback, input-feedthrough, and gains. We combine them and consider a transformed plant $\tilde{\Sigma}$ with new input $\tilde{u}$ and output $\tilde{y}$, which are connected to $u,y$ via 
\begin{align}\label{eq.Transformation}
\left[\begin{smallmatrix}\tilde{u} \\ \tilde{y}\end{smallmatrix}\right] = T \left[\begin{smallmatrix} u \\ y\end{smallmatrix}\right],
\end{align}
for some invertible matrix $T$. The action of the transformation $T$ on the a system $\Sigma$ can be seen in the block diagram in Fig. \ref{fig.BlockDiagramT}. This transformation is an aggregation of a constant gain input-feedthrough, constant gain output-feedback, and cascade with a constant gain, as seen in Fig. \ref{fig.BlockDiagram} (see \cite{Sharf2019a} for further details). 
We wish to understand the effect of these I/O transformations on the passivity of the transformed system. Formally, the problem we consider is the following:

\begin{figure}[t!]
\centering
\includegraphics[width=5.7cm]{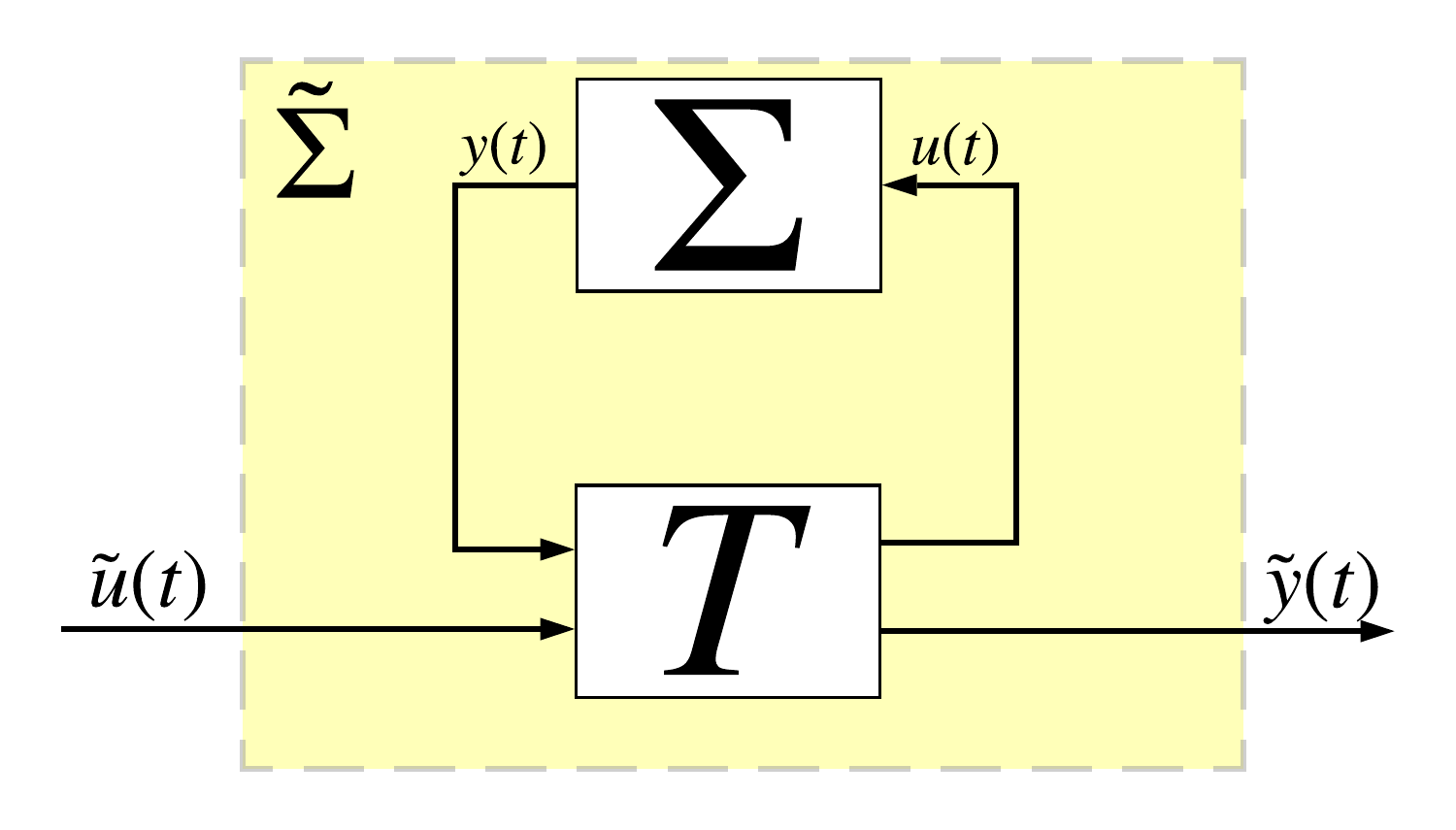}
\caption{A block diagram describing the transformed system $\tilde{\Sigma}$ after the linear transformation $T$.}
\label{fig.BlockDiagramT}
\end{figure} 

\begin{problem} \label{prob.Passivize}
Let $\Sigma$ be a dynamical system with equal input and output dimensions, which is I/O $(\rho,\nu)$-passive, and let $\rho_\star,\nu_\star$ be numbers such that $\rho_\star\nu_\star < 1/4$. Characterize all I/O transformations of the form \eqref{eq.Transformation} such that the transformed system $\tilde{\Sigma}$ is I/O $(\rho_\star,\nu_\star)$-passive.
\end{problem}

In order to address this problem, we consider the geometric approach to passivity of \cite{Sharf2019a}. It considers an abstraction of the inequalities appearing in Definition \ref{def.WPassivity}, called projective quadratic inequalities:

\begin{figure}[t!]
\centering
\includegraphics[width=7.7cm]{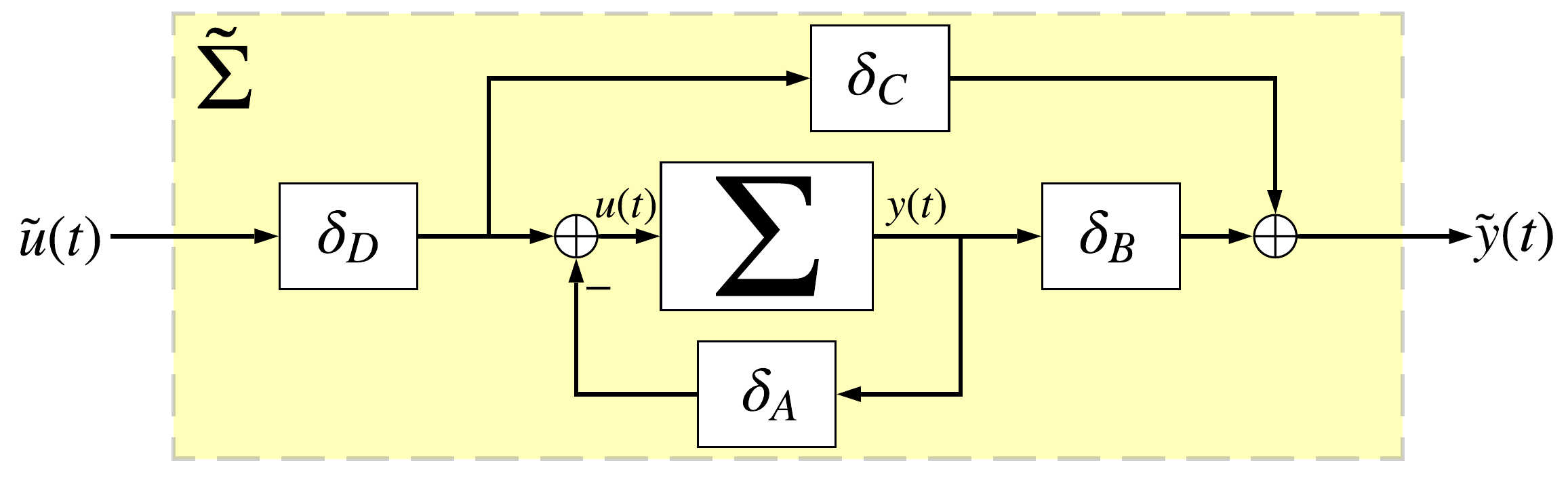}
\caption[The transformed system $\tilde{\Sigma}$ after the linear transformation $T$.]{A different block diagram describing the transformed system $\tilde{\Sigma}$ after the linear transformation $T$, as in Fig. \ref{fig.BlockDiagramT}. If $\Sigma$ is SISO and $T=\left[\protect\begin{smallmatrix} a & b \\ c & d \protect\end{smallmatrix}\right]$, then $\delta_A = b/a, \delta_B =d-\frac{b}{a}c,  \delta_C = c$ and $\delta_D = a$.} 
\label{fig.BlockDiagram}
\end{figure}

\begin{definition}[PQI]
A $d$-dimensional \emph{projective quadratic inequality} (PQI) is an inequality in the variables $\xi,\chi \in\R^d$ of the form
\begin{align}
0 \le a \|\xi\|^2 + b\xi^\top \chi + c\|\chi\|^2 :=\mathrm{\bf f}_{(a,b,c)}(\xi,\chi),
\end{align}
for some numbers $a,b,c$, not all zero. The inequality is called \emph{non-trivial} if $b^2-4ac>0$. The associated solution set of the $d$-dimensional PQI is the set of all points $(\xi,\chi)\in\R^{2d}$ satisfying the inequality. If $d=1$, we'll omit the dimension and call the inequality a PQI.
\end{definition}

\textcolor{black}{PQIs are a special case of the sector bound formulation for passivity, see e.g. \cite{Safonov1982}}.
We note that $d$-dimensional PQI resembles \eqref{eq.IOWPassivity}, in which $\xi,\chi,a,b,c$ are replaced by $u,y,-\nu,1,-\rho$ respectively. For this reason, we denote $\mathrm{\bf f}_{(a,b,c)}(\xi,\chi)$ as $\varphi_{\rho,\nu}(\xi,\chi)$, and the corresponding solution set as $C_{\rho,\nu,d}$. Namely, we define $\varphi_{\rho,\nu}$ as 
$\varphi_{\rho,\nu}(\xi,\chi)=-\nu\|\xi\|^2+\xi^\top\chi -\rho\|\chi\|^2$ 
and $C_{\rho,\nu,d}$ as $\{(\xi,\chi)\in \mathbb{R}^d\times\mathbb{R}^d:\ \varphi_{\rho,\nu}(\xi,\chi) \ge 0\}$. For convenience, we denote the latter as $C_{\rho,\nu}$ when $d=1$.

\begin{remark}
The definition of $d$-dimensional PQIs allows an abstraction of the inequality defining passivity. It also encapsulates more sophisticated variants of passivity, such as shifted passivity, incremental passivity [\cite{Pavlov2008}], equilibrium-independent passivity [\cite{Hines2011}] and maximal equilibrium-independent passivity [\cite{Burger2014,Sharf2018a}]. Hence, the results of the paper also apply to these variants. 
\end{remark}

As noted in \cite{Sharf2019a}, I/O transformations give rise to an action of the group of $2\times 2$ invertible matrices, $GL_{2}(\R)$, on the collection of solution sets of PQIs. \textcolor{black}{This allows us to use standard group theory methods, which are manifested e.g. in Proposition \ref{proposition.3Prod}}
In particular, let $A$ be the solution set of a ($d$-dimensional) PQI, $A=\{(\xi,\chi)\,:\,0\le \mathrm{\bf f}_{(a,b,c)}(\xi,\chi)\}$. For any invertible matrix $T\in GL_{2d}(\R)$, the solution set of the transformed ($d$-dimensional) PQI is given by $T(A)$, the image of $A$ under $T$.
In fact, one can show that an I/O transformation maps an I/O $(\rho,\nu)$-passive system to an I/O $(\rho_\star,\nu_\star)$-passive system if and only if it maps the $d$-dimensional PQI $0\le \varphi_{\rho,\nu}(\xi,\chi)$ to the $d$-dimensional PQI $0\le \varphi_{\rho_\star,\nu_\star}(\xi,\chi)$ (or to a stricter inequality).

Following \cite{Sharf2019a}, we first focus on the case of SISO systems. 

\begin{definition}
A \emph{symmetric section} $S$ on the unit circle $\BS^1 \subseteq \R^2$ is defined as the union of two closed disjoint sections that are opposite to each other, i.e., $S=B\cup(-B)$, where $B$ is a closed section of angle $<\pi$. A \emph{symmetric double cone} is defined as $A=\{\lambda s:\ \lambda >0 ,s\in S\}$ for some symmetric section $S$.
\end{definition}

The connection between cones and passivity theory is intricate, stemming from the notion of sector-bounded nonlinearities [\cite{Zames1966,McCourt2009}].  An example of a symmetric section and the associated symmetric double-cone can be seen in Fig. \ref{fig.SymmetricSection}. These are of interest due to their close relationship with (1-dimensional) PQIs. Namely,
\begin{theorem}[\cite{Sharf2019a}]
The solution set of any non-trivial PQI is a symmetric double cone. Moreover, any symmetric double-cone is the solution set of some non-trivial PQI, which is unique up to a multiplicative positive constant.
\end{theorem}
As a corollary, we conclude that a map transforms an I/O $(\rho,\nu)$-passive system to an I/O $(\rho_\star,\nu_\star)$-passive system if and only if it sends $C_{\rho,\nu}$ into $C_{\rho_\star,\nu_\star}$, which we denote by $C_{\rho,\nu} \hookrightarrow  C_{\rho_\star,\nu_\star}$  Thus, we wish to characterize maps $C_{\rho,\nu} \hookrightarrow C_{\rho_\star,\nu_\star}$. One possible transformation is given in the following theorem:

\begin{figure}[b!]
\centering
\includegraphics[width=3.7cm]{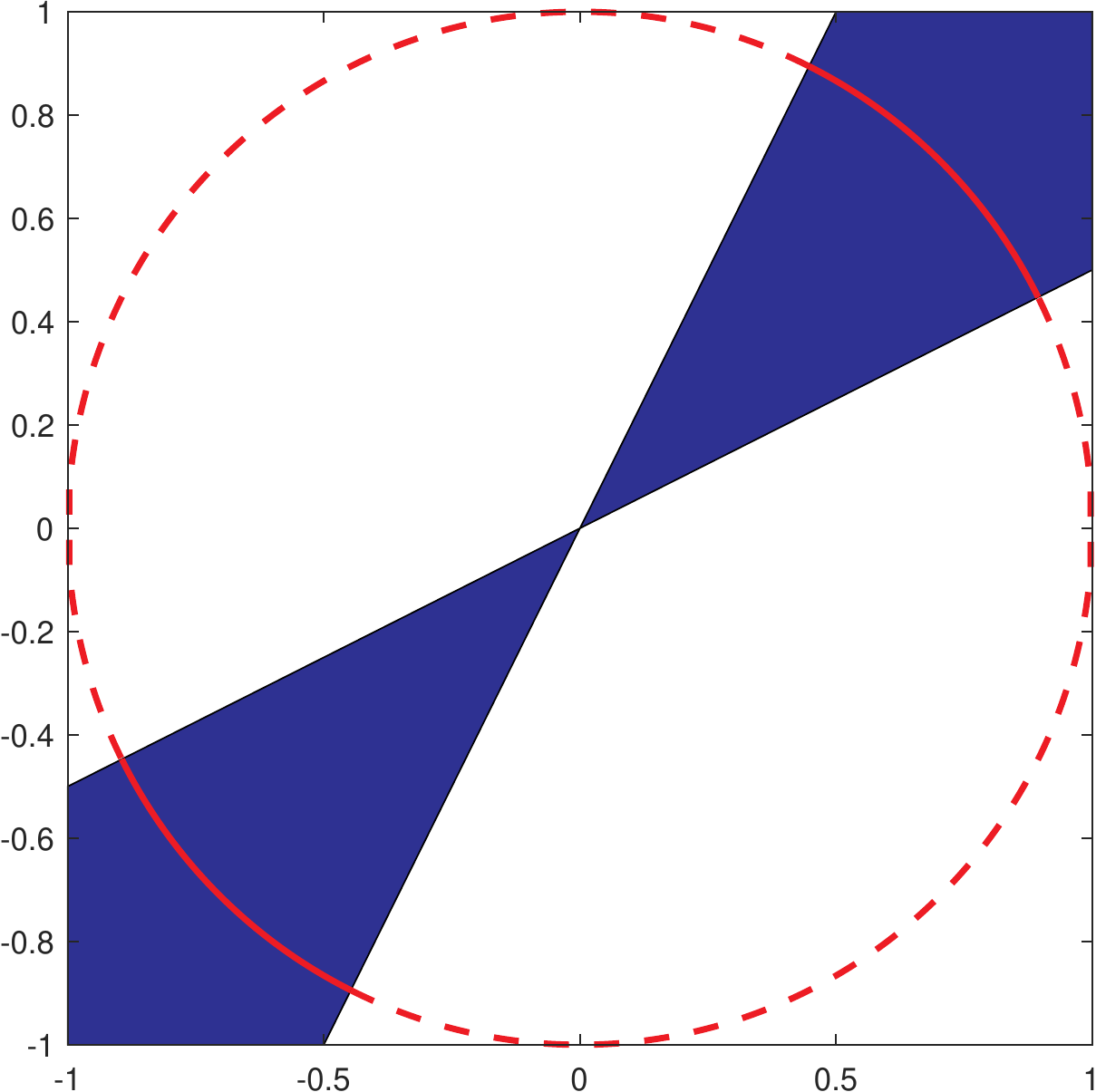}
\caption{A double cone (in blue), and the associated symmetric section (in solid red). The parts of $\mathbb{S}^1$ outside the symmetric section are presented by the dashed red line.} 
\label{fig.SymmetricSection}
\end{figure} 

\begin{theorem}[\cite{Sharf2019a}] \label{theorem.MappingConesIntoCones}
Let $\rho$,$\nu$,$\rho_\star,\nu_\star$ be any numbers such that $\rho\nu,\rho_\star\nu_\star < 1/4$. Let $(\xi_1,\chi_1)$ and $(\xi_2,\chi_2)$ be two non-colinear solutions to $\varphi_{\rho,\nu}(\xi,\chi)= 0$. Moreover, let $(\xi_3,\chi_3)$ and $(\xi_4,\chi_4)$ be two non-colinear solutions to $\varphi_{\rho_\star,\nu_\star}(\xi,\chi) = 0$. Define
\begin{align*}
T_1=\left[\begin{matrix}\xi_3 & \xi_4 \\ \chi_3 & \chi_4\end{matrix}\right]\left[ \begin{matrix}\xi_1 & \xi_2 \\ \chi_1 & \chi_2\end{matrix}\right]^{-1}, T_2=\left[\begin{matrix}\xi_3 & -\xi_4 \\ \chi_3 & -\chi_4\end{matrix}\right]\left[ \begin{matrix}\xi_1 & \xi_2 \\ \chi_1 & \chi_2\end{matrix}\right]^{-1}.
\end{align*}
Let $\alpha_1$ be equal to $1$ if $\varphi_{\rho,\nu}(\xi_1+\xi_2,\chi_1+\chi_2) \ge 0$ and zero otherwise. Moreover let $\alpha_2$ be equal to $1$ if $\varphi_{\rho_\star,\nu_\star}(\xi_3+\xi_4,\chi_3+\chi_4) \ge 0$ and zero otherwise.
\begin{itemize}
\item[i)] If $\alpha_1 = \alpha_2$, then $T_1$ is $C_{\rho,\nu}\hookrightarrow C_{\rho_\star,\nu_\star}$.
\item[ii)] If $\alpha_1 \neq \alpha_2$, then $T_2$ is $C_{\rho,\nu}\hookrightarrow C_{\rho_\star,\nu_\star}$.
\end{itemize}
\end{theorem}

\textcolor{black}{
Before moving forward characterizing all passivizing transformations, we wish to clarify their applicability:}
\textcolor{black}{
\begin{rem}
Fig. \ref{fig.BlockDiagram} shows that any transformation $T$ of the form \eqref{eq.Transformation} can be applied using four basic blocks. Using block diagram manipulation, we can show the feedback connection of a transformed system $\tilde{\Sigma}$ with a controller $C$ is equivalent to the feedback connection of the original system $\Sigma$ with a transformed controller $\tilde{C}$. 
\end{rem}}
\textcolor{black}{
Moreover, these transformations allow one to extend standard passivity-based arguments to I/O $(\rho,\nu)$-passive systems. }
\textcolor{black}{
\begin{proposition}
Consider a system $\Sigma$ with input $u(t)$ and output $y(t)$. Suppose that the transformation $T \in GL_{2n}(\R)$ maps the system $\Sigma$ to an output-strictly passive system $\tilde{\Sigma}$, and consider the feedback connection of $\tilde{\Sigma}$ with an output-strictly passive controller $C$. Then the output $y(t)$ is regulated to $0$ as $t\to \infty$.
\end{proposition}}
\vspace{-40pt}
\textcolor{black}{
\begin{pf}
LaSalle's invariance principle shows that the transformed input $\tilde{u}(t)$ and the transformed output $\tilde{y}(t)$ both converge to zero. The invertability of $T$, together with \eqref{eq.Transformation}, imply that $y(t) \to 0$.
\end{pf}
In other words, regulation of the output $y(t)$ of the system $\Sigma$ can be achieved by regulating  the output of the transformed system $\tilde{\Sigma}$ with output $\tilde{y}$.
}
\vspace{-15pt}
\section{A Characterization of All Passivizing Transformations for SISO Systems} \label{sec.SISO}
\vspace{-10pt}
We wish to characterize all I/O transformations mapping an arbitrary dynamical system $\Sigma$ to an I/O $(\rho_\star,\nu_\star)$-passive system. Namely, we assume that the given system is I/O $(\rho,\nu)$-passive (for some known $\rho,\nu$), and seek all transformations that force the transformed system to be I/O $(\rho_\star,\nu_\star)$-passive. We do so by finding all transformations that map a given double cone $C_{\rho,\nu}$ into $C_{\rho_\star,\nu_\star}$. Theorem \ref{theorem.MappingConesIntoCones} provides one way to build a map from an arbitrary cone into another arbitrary cone, but does not prescribe a general method to find all such maps. However, we can use Theorem \ref{theorem.MappingConesIntoCones} to show that all maps from an arbitrary cone into another arbitrary cone can be built using maps from $C_{0,0}$ into itself.
\begin{proposition} \label{proposition.3Prod}
Let $\rho,\nu,\rho_\star,\nu_\star$ be any four numbers such that $\rho\nu,\rho_\star\nu_\star < 1/4$, and let $T$ be any matrix $C_{\rho,\nu} \hookrightarrow C_{\rho_\star,\nu_\star}$. Let $S_{\rho,\nu},S_{\rho_\star,\nu_\star}$ be the invertible matrices $C_{0,0} \hookrightarrow C_{\rho,\nu},  C_{0,0} \hookrightarrow C_{\rho_\star,\nu_\star}$ respectively, as built using Theorem \ref{theorem.MappingConesIntoCones}. Then there exists a matrix $Q$, which is $C_{0,0}\hookrightarrow C_{0,0}$, such that $T = S_{\rho_\star,\nu_\star}QS_{\rho,\nu}^{-1}$ holds.
\end{proposition}
\vspace{-15pt}
\begin{pf}
Theorem \ref{theorem.MappingConesIntoCones} shows that $S^{-1}_{\rho,\nu},S^{-1}_{\rho_\star,\nu_\star}$ map  $C_{\rho,\nu}$ and $C_{\rho_\star,\nu_\star}$ into $C_{0,0}$, respectively. Define $Q=S_{\rho_\star,\nu_\star}^{-1}TS_{\rho,\nu}$. Then $Q$ is invertible as a product of invertible matrices. Moreover, it maps $C_{0,0}$ into itself as:
\begin{align*}
C_{0,0} \overset{S_{\rho,\nu}}{\hookrightarrow}C_{\rho,\nu}\overset{T}{\hookrightarrow}C_{\rho_\star,\nu_\star}\overset{S_{\rho_\star,\nu_\star}^{-1}}{\hookrightarrow}C_{0,0}.
\end{align*}\demo
\end{pf}
Proposition \ref{proposition.3Prod} gives a prescription for finding all matrices mapping $C_{\rho,\nu}$ into $C_{\rho_\star,\nu_\star}$. It contains two main ingredients, namely the matrices $S_{\mu,\tau}$, and matrices mapping $C_{0,0}$ into itself. We start by finding all matrices in $GL_{2}(\R)$ mapping $C_{0,0}$ into itself:

\begin{proposition} \label{pro.C00Transformation}
A matrix $T\in GL_2(\R)$ sends $C_{0,0}$ into itself if and only if all of the entries of $T$ have the same sign, i.e. $T_{ij}T_{kl} \ge 0$ for every $i,j,k,l\in\{1,2\}$.
\end{proposition}

\begin{pf}
We first show that if $T$ sends $C_{0,0}$ into itself, then all of the entries of $T=(T_{ij})_{i,j}$ have the same sign. We recall that $C_{0,0}$ contains all points $(\xi,\chi)$ such that $\xi\chi \ge 0$, i.e., $C_{0,0}$ is a union of $\{0\}$, the first quadrant, and the third quadrant. We note that $e_1 = (1,0)^\top$ and $e_2 = (0,1)^\top$ are in $C_{0,0}$, hence $Te_1 = (T_{11},T_{21})^\top$ and $Te_2 = (T_{12},T_{22})^\top$ are also in $C_{0,0}$. This implies that $T_{11},T_{21}$ have the same sign, and that $T_{12},T_{22}$ have the same sign, and in each pair not both elements are zero (as $T$ is invertible). We note that by switching between $T$ and $-T$, we may assume without loss of generality that $T_{11},T_{21}$ are both non-negative. We want to show that $T_{12},T_{22}$ are also both non-negative.

Assume the contrary, that is, that $T_{12},T_{22}$ are both non-positive. Moreover, as $Te_1,Te_2 \neq 0$, we conclude that $Te_1$ lies in the first quadrant of $\R^2$, and that $Te_2$ lies in the third quadrant. We note that the line between $e_1,e_2$ lies inside $C_{0,0}$, so the same is true for the line between $Te_1,Te_2$, as $T$ is linear and maps $C_{0,0}$ into itself. However, as $Te_1$ is in the first quadrant and $Te_2$ is in the third, the straight line between them passes either through zero, the second quadrant or the fourth quadrant. The latter two cases are impossible, as $C_{0,0}$ contains no points from these quadrants, and the former case is impossible as it would imply that the invertible transformation $T$ maps a non-zero point to zero. We thus conclude all entries of $T$ have the same sign.

Conversely, assume that all of the entries of $T$ have the same sign. By replacing $-T$ with $T$, we assume without loss of generality that $T_{ij} \ge 0$ for all $i,j\in\{1,2\}$, so that $Te_1,Te_2$ are both in the first quadrant. Take any point $x\in C_{0,0}$. If $x=0$ then $Tx = 0\in C_{0,0}$. If $x$ is in the first quadrant, then it is a linear combination of $e_1,e_2$ with non-negative coefficients, not both zero. Thus $Tx$ is a linear combination of $Te_1,Te_2$ with non-negative coefficients (not both zero), hence $Tx$ is in the first quadrant. If $x$ is in the third quadrant, then $-x$ is in the first quadrant, so $T(-x)=-Tx$ is in the first quadrant, hence $Tx$ is in the third quadrant. As we showed that $Tx \in C_{0,0}$ for all $x \in C_{0,0}$, this concludes the proof.\demo
\end{pf}

\begin{rem}
More generally, given some $\rho,\nu$, one could ask for a characterization of all matrices $T\in GL_2(\R)$ that are $C_{\rho,\nu} \hookrightarrow C_{\rho,\nu}$. Mimicking the proof above, one can show that a map $T\in GL_2(\R)$ is $C_{\rho,\nu} \hookrightarrow C_{\rho,\nu}$ if and only if all of the elements of the matrix $I_{e\to\B} T I_{e\to\B}^{-1}$ possess the same sign, where $\B$ is composed of the non-colinear solutions to the equation $-\nu \xi^2 + \xi\chi - \rho\chi^2=0$ and $e$ is the standard basis. A more explicit form for the basis $\B$ can be achieved by taking the columns of the matrix $S_{\rho,\nu}$, as seen in Proposition \ref{proposition.ExplicitConnectingMatrices} below.
\end{rem}

We now clarify the second component appearing in Proposition \ref{proposition.3Prod}, namely the matrices $S_{\mu,\tau}$:
\begin{proposition}\label{proposition.ExplicitConnectingMatrices}
Let $\mu,\tau$ be any two numbers such that $\mu\tau < 1/4$. Recall that $S_{\mu,\tau}$ is a map $C_{0,0}\hookrightarrow C_{\mu,\tau}$, as constructed in Theorem \ref{theorem.MappingConesIntoCones}. Define $R = \sqrt{1-4\tau\mu}$.
\begin{itemize}
\item [i)] If $\tau < 0$, we can choose 
$
S_{\mu,\tau} = \frac{1}{2\tau}\left[\begin{smallmatrix} -1 - R  & 1 - R \\ -2\tau  & 2\tau \end{smallmatrix}\right].
$\\
\item [ii)] If $\tau > 0$,, we can choose 
$
S_{\mu,\tau} = \frac{1}{2\tau}\left[\begin{smallmatrix} 1 + R  & 1 - R \\ 2\tau  & 2\tau \end{smallmatrix}\right].
$\\
\item [iii)] If $\tau = 0$, we can choose 
$
S_{\mu,\tau} = \left[\begin{smallmatrix} 1 & \mu \\ 0 & 1 \end{smallmatrix}\right].
$
\end{itemize} 
\end{proposition}
\vspace{-10pt}
\begin{pf}
We use Theorem \ref{theorem.MappingConesIntoCones} to build $S_{\mu,\tau}$. As we consider a map $C_{0,0}\hookrightarrow C_{\mu,\tau}$, we take $(\xi_1,\chi_1)=(1,0)$ and $(\xi_2,\chi_2) = (0,1)$. As $(\xi_1+\xi_2,\chi_1+\chi_2)=(1,1)$ satisfies the PQI $\xi\chi \ge 0$, we choose:
\begin{align*}
S_{\mu,\tau} = \begin{cases} \begin{bmatrix} \xi_3 & \xi_4 \\ \chi_3 & \chi_4 \end{bmatrix} & \alpha_2 = 1  \\ 
\begin{bmatrix} \xi_3 & -\xi_4 \\ \chi_3 & -\chi_4 \end{bmatrix} & \alpha_2 \neq 1  \\\end{cases},
\end{align*}
where we recall that $(\xi_3,\chi_3), (\xi_4,\chi_4)$ are two non-colinear solutions to $-\tau \xi^2 + \xi\chi - \mu \chi^2 = 0$, and $\alpha_2 = 1$ if and only if $(\xi_3+\xi_4,\chi_3+\chi_4)$ satisfies the PQI $-\tau \xi^2 + \xi\chi - \mu \chi^2 \ge 0$. 
We first assume that $\tau \neq 0$, so we write $-\tau \xi^2 + \xi\chi - \mu \chi^2 = 0$ as $-\tau(\xi-a_1\chi)(\xi-a_2\chi)=0$, where $a_1,a_2$ are given by 
\begin{align*}
a_1 =& \frac{-1+\sqrt{1-4\tau\mu}}{-2\tau} = \frac{1- R}{2\tau} \\
a_2 =& \frac{-1-\sqrt{1-4\tau\mu}}{-2\tau} = \frac{1+ R}{2\tau},
\end{align*}
where we note that $a_1\neq a_2$ as $\mu\tau < 1/4$. Choose $(\xi_3,\chi_3) = (-a_2,-1), (\xi_4,\chi_4) = (a_1,1)$. We have that the point $(\xi_3+\chi_3,\xi_4+\chi_4) = (a_1-a_2,0)$ satisfies the PQI $-\tau \xi^2 + \xi\chi - \mu \chi^2 \ge 0$ if and only if $\tau <0$, as we assumed $\tau \neq 0$. We therefore conclude the desired result for $\tau \neq 0$ from Theorem \ref{theorem.MappingConesIntoCones}. 

Suppose now that $\tau = 0$. We note that $(\xi_3,\chi_3) = (1,0)$ and $(\xi_4,\chi_4) = (\mu,1)$ are two non-colinear solutions to $\xi\chi - \mu \chi^2 = 0$, and that $(\xi_3+\xi_4,\chi_3+\chi_4) = (1+\mu,1)$ satisfies the PQI $-\tau\xi^2 + \xi\chi - \mu \chi^2 \ge 0$, as $\xi\chi - \mu\chi^2 = \chi(\xi - \mu \chi)$. This completes the proof.\demo
\end{pf}

We now conclude with the following theorem:
\begin{theorem} \label{theorem.SISOProduct}
Let $\Sigma$ be a SISO I/O $(\rho,\nu)$-passive system, and let $T \in GL_2(\R)$ be an invertible matrix inducing an I/O transformation of the form \eqref{eq.Transformation}. The transformed system $\tilde{\Sigma}$ is I/O $(\rho_\star,\nu_\star)$-passive if and only if there exists a matrix $M\in GL_2(\R)$ such that $M_{ij} \ge 0$ for all $i,j\in\{1,2\}$ and some $\theta \in \{\pm 1\}$ such that $T = \theta S_{\rho_\star,\nu_\star}MS_{\rho,\nu}^{-1}$, where $S_{\rho,\nu},S_{\rho_\star,\nu_\star}$ are given in Proposition \ref{proposition.ExplicitConnectingMatrices}. In other words, the transformed system $\tilde{\Sigma}$ is I/O $(\rho_\star,\nu_\star)$-passive if and only if all of the entries of the matrix $ S_{\rho_\star,\nu_\star}^{-1}TS_{\rho,\nu}$ have the same sign.
\end{theorem}

\begin{figure}[b!]
\centering
\includegraphics[width=7.7cm]{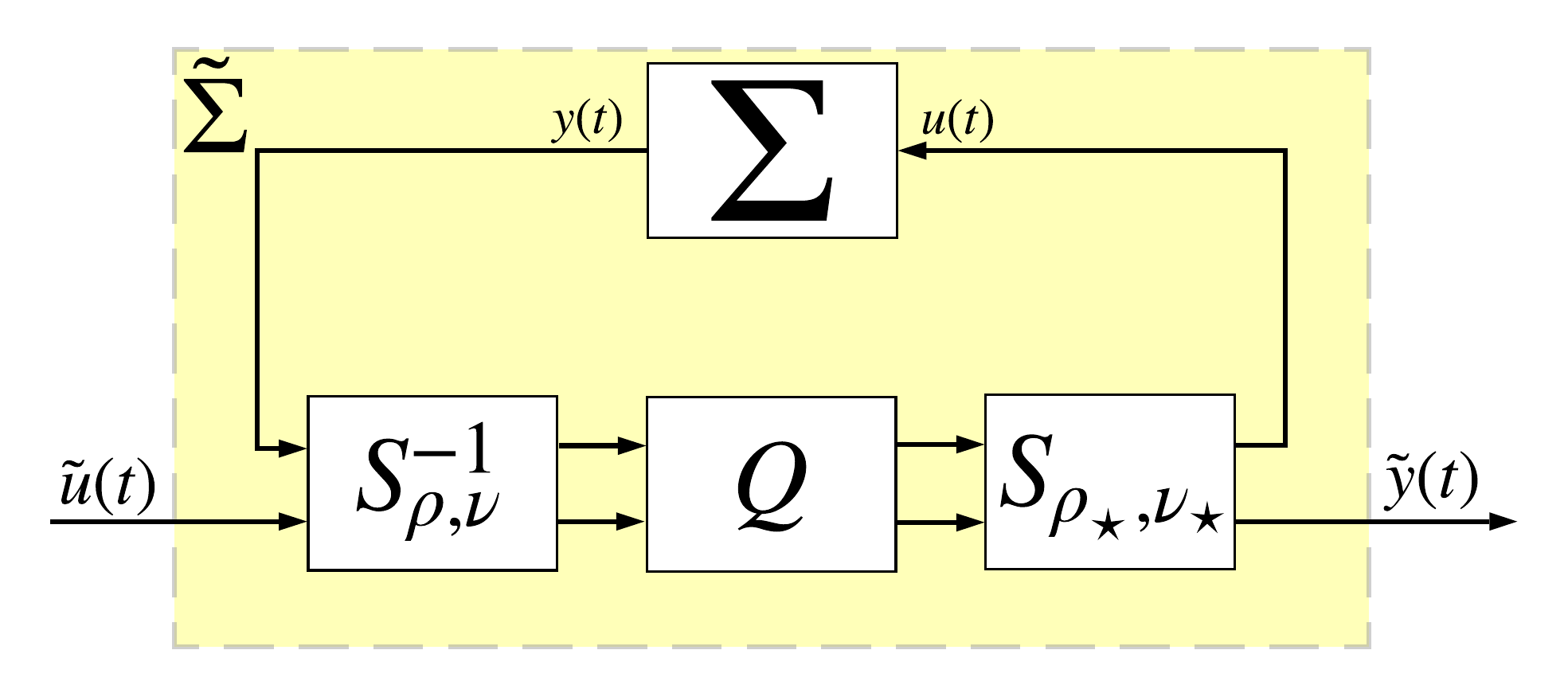}
\caption{A general transformation mapping a SISO I/O $(\rho,\nu)$-passive system to a SISO I/O $(\rho_\star,\nu_\star)$-passive system. The entries of the matrix $Q\in\mathbb{R}^{2\times 2}$ all have the same sign.}
\label{fig.BlockDiagramTPassivize}
\end{figure}

A block diagram visualizing Theorem \ref{theorem.SISOProduct} can be seen in Fig. \ref{fig.BlockDiagramTPassivize}.

\begin{pf}
Proposition \ref{proposition.3Prod} implies that for an invertible matrix $T \in GL_2(\R)$, the transformed system $\tilde{\Sigma}$ is I/O $(\rho_\star,\nu_\star)$-passive if and only if there exists an invertible matrix $Q\in GL_2(\R)$ which is $C_{0,0} \hookrightarrow C_{0,0}$ such that $T = S_{\rho_\star,\nu_\star}QS_{\rho,\nu}^{-1}$. By Proposition \ref{pro.C00Transformation}, a matrix $Q$ is $C_{0,0} \hookrightarrow C_{0,0}$ if and only if all of its entries possess the same sign. By letting $\theta \in \{\pm 1\}$ be that sign, we can write any matrix $Q$ sending $C_{0,0}$ into itself as $Q = \theta M$, where $M\in GL_2 (\R)$ and $M_{ij} \ge 0$ for all $i,j\in \{1,2\}$. Thus, the transformed system $\tilde{\Sigma}$ is I/O $(\rho_\star,\nu_\star)$-passive if and only if there exists some $\theta \in \{\pm 1\}$ and $M\in GL_2(\R)$ with non-negative entries such that $T = \theta S_{\rho_\star,\nu_\star}MS_{\rho,\nu}^{-1}$.\demo
\end{pf}

\section{Extension to MIMO Systems} \label{sec.MIMO}
Up to now, we gave an explicit description of all I/O transformations mapping I/O $(\rho,\nu)$-passive \emph{SISO} systems to I/O $(\rho_\star,\nu_\star)$-passive \emph{SISO} systems. One could try and generalize this idea to MIMO systems, but a few problems arise. The cornerstone in the characterization for SISO systems was Theorem \ref{theorem.MappingConesIntoCones}, whose proof uses the fact that for SISO systems, the solution sets of PQIs are two-dimensional, and their boundary is the union of two straight lines [\cite{Sharf2019a}]. For $d\times d$ MIMO systems, the solution set of a PQI lies in $\R^{2d}$, and its boundary, in general, is of dimension $2d-1$ (almost everywhere). Thus, a geometric framework for MIMO systems cannot be applied easily.

To deal with the MIMO case, we use a similar idea, studying the action of the collection of $2d\times 2d$ invertible linear transformations, $GL_{2d}(\R)$, on the collection of $d$-dimensional PQIs. As before, we use the notion of solution sets. We recall that we denoted the solution set of the $d$-dimensional PQI $\varphi_{\rho,\nu}(\xi,\chi) \ge 0$ by $C_{\rho,\nu,d}$. As before, $T$ maps one $d$-dimensional PQI to another if and only if it maps the associated solution sets to the another. We start with the following proposition:
\begin{proposition} \label{proposition.Kronecker}
Let $\rho,\nu,\rho_\star,\nu_\star$ be any real numbers, and let $S\in GL_{2}(\R)$ be any matrix mapping the 1-dimensional PQI $0\le -\nu\xi^2 + \xi \chi - \rho \chi^2$ to the 1-dimensional PQI $0\le -\nu_\star\xi^2 + \xi \chi - \rho_\star \chi^2$. Then $S\otimes {\rm Id}_d$ maps the $d$-dimensional PQI $-\nu\|\xi\|^2 + \xi^\top \chi - \rho \|\chi\|^2$ to the $d$-dimensional PQI $-\nu_\star\|\xi\|^2 + \xi^\top \chi - \rho_\star \|\chi\|^2$.
\end{proposition}

Thus, the MIMO analogue of the transformations $S_{\rho,\nu}$ are $S_{\rho,\nu}\otimes{\rm Id}_d$. We now prove the proposition.

\begin{pf} We define 
$
A = \left[\begin{smallmatrix}-\rho & 0.5 \\ 0.5 & -\nu\end{smallmatrix}\right], B = \left[\begin{smallmatrix}-\rho_\star & 0.5 \\ 0.5 & -\nu_\star\end{smallmatrix}\right].
$
The 1-dimensional PQI $0\le -\nu_\star\xi^2 + \xi \chi - \rho_\star \chi^2$ can be written as $\Xi^\top A \Xi\ge 0$, where $\Xi = [\chi, \xi]^\top\in\R^2$, and the 1-dimensional PQI $0\le -\nu_\star\xi^2 + \xi \chi - \rho_\star \chi^2$ is written as $\Xi^\top B \Xi\ge 0$. By setting $\tilde{\Xi} = S\Xi$, we see that $S$ maps the first 1-dimensional PQI to the second if and only if $(S^{-1})^\top A S^{-1} = B$, and the latter condition implies 
$$
((S\otimes{\rm Id}_d)^{-1})^\top (A\otimes{\rm Id}_d)(S\otimes{\rm Id}_d)^{-1} = B\otimes{\rm Id}_d.
$$ 
The proof in now complete, as we note the $d$-dimensional PQIs can be written as $\Xi_d^\top (A\otimes{\rm Id}_d)\Xi_d \ge 0$ and $\Xi_d^\top (B\otimes {\rm Id}_d)\Xi_d \ge 0$, where $\Xi_d = [\xi^\top,\chi^\top]^\top\in\R^{2d}$.\demo
\end{pf}

\begin{rem}
Proposition \ref{proposition.Kronecker} does \emph{not} claim that \emph{all} maps between $d$-dimensional PQIs stem from maps between $1$-dimensional PQIs using the Kronecker product.
\end{rem}

We now search for a MIMO analogue for the second component we had, namely non-negative matrices. Before, non-negative matrices stemmed from maps $C_{0,0}\hookrightarrow C_{0,0}$. 
\begin{proposition} \label{proposition.Riccati}
An invertible matrix $T\in GL_{2d}(\R)$ maps $C_{0,0,d}$ into itself if and only if there exists some $\lambda > 0$ such that $T^\top J T - \lambda J \ge 0$, where 
$
J = \left[\begin{smallmatrix} 0  & 0.5{\rm Id}_d \\ 0.5{\rm Id}_d & 0 \end{smallmatrix}\right].
$
\end{proposition}
\begin{pf}
As before, we denote the stacked variable vector as $\Xi_d = [\xi^\top,\chi^\top]^\top\in \R^{2d}$. The set $C_{0,0,d}$ is the collection of all vectors $\Xi_d$ satisfying $\Xi_d^\top J \Xi_d \ge 0$. The image of $C_{0,0,d}$ under $T$ consists of all vectors $\tilde{\Xi}_d$ such that $\tilde{\Xi}_d^\top (T^{-1})^\top J T^{-1}\tilde{\Xi}_d \ge 0$. Thus, $T$ maps $C_{0,0,d}$ inside itself if and only if the following implication holds:
\begin{align*}
\tilde{\Xi}_d^\top (T^{-1})^\top J T^{-1}\tilde{\Xi}_d \ge 0 \implies \tilde{\Xi}_d^\top J \tilde{\Xi}_d \ge 0,~~~\forall \tilde{\Xi}_d\in\R^{2d}.
\end{align*}
By the S-lemma, or S-procedure, [\cite[Appendix B]{Boyd2004}], the above implication is equivalent to the existence of some $\mu > 0$ such that $(T^{-1})^\top J T^{-1} - \mu J \le 0$. By multiplying the inequality by $T^\top$ on the left and by $\mu^{-1} T$ on the right, the inequality is equivalent to $T^\top J T - \lambda J \ge 0$, where $\lambda = \mu^{-1} > 0$.\demo
\end{pf}

\begin{rem}
The inequality $T^\top J T - \lambda J \ge 0$ can be seen as a certain generalized version of an algebraic Riccati inequality. Indeed, the algebraic Riccati equation is given as $A^\top P + PA - PXP + Q= 0$, where $X,Q$ are positive-definite matrices, and $P$ is a symmetric matrix variable. The associated inequality, $A^\top P + PA - PXP + Q \le 0$, has also been considered in literature [\cite{Willems1971}]. Choosing $Q=\lambda J, A=0$ and $X = J$, and not restricting the matrix $P$ to be symmetric, results in the inequality $P^\top J P - \lambda J \ge 0$.\footnote{We have to use $P^\top X P$ instead of $PXP$ to guarantee that the matrix is symmetric.} As $Q,J$ are not positive definite, this is a generalized version of an algebraic Riccati inequality.
\end{rem}

Combining Propositions \ref{proposition.Kronecker} and \ref{proposition.Riccati}, we conclude with the following theorem:
\begin{theorem} \label{thm.MIMOCharacter}
Let $\Sigma$ be an I/O $(\rho,\nu)$-passive system with input and output dimension equal to $d$, and let $T \in GL_{2d}(\R)$ be an invertible matrix inducing an I/O transformation of the form \eqref{eq.Transformation}. The transformed system $\tilde{\Sigma}$ is I/O $(\rho_\star,\nu_\star)$-passive if and only if there exists a matrix $M \in GL_{2d}(\R)$ and some positive $\lambda > 0$ such that:
\begin{align*}
T = (S_{\rho_\star,\nu_\star} \otimes {\rm Id}_d)M(S_{\rho,\nu}^{-1} \otimes {\rm Id}_d),\ \  
M^\top J M - \lambda J \ge 0,
\end{align*}
where 
$
J = \left[\begin{smallmatrix} 0  & 0.5{\rm Id}_d \\ 0.5{\rm Id}_d & 0 \end{smallmatrix}\right]
$,
i.e., $\tilde{\Sigma}$ is I/O $(\rho_\star,\nu_\star)$-passive if and only if there exists $\lambda > 0$ such that  $X = (S_{\rho_\star,\nu_\star}^{-1} \otimes {\rm Id}_d)T(S_{\rho,\nu} \otimes {\rm Id}_d)$ satisfies $X^\top J X - \lambda J \ge 0$.
\end{theorem}

\begin{pf}
The proof is nearly identical to that of Proposition \ref{proposition.3Prod}, where we replace the sets $C_{\rho,\nu},C_{\rho_\star,\nu_\star}$ by the corresponding $d$-dimensional PQIs.
\demo
\end{pf}

The theorem can be seen as a generalization of Theorem \ref{theorem.SISOProduct}, as one can verify that for $d=1$, $X^\top J X - \lambda J \ge 0$ for some $\lambda > 0$ if and only if all of $X$-s entries possess the same sign. Indeed,

\begin{proposition} \label{prop.SameSign}
Let $X \in GL_2(\R)$, and let $J = \left[\begin{smallmatrix} 0  & 0.5 \\ 0.5 & 0 \end{smallmatrix}\right]$.  There exists some $\lambda > 0$ such that $X^\top J X - \lambda J \ge 0$ if and only if all of the entries of $X$ possess the same sign.
\end{proposition}
\begin{pf}
Write $X = \left[\begin{smallmatrix} a & b \\ c & d \end{smallmatrix}\right]$. The matrix $X^\top J X - \lambda J$ can be computed as:
\begin{align*}
X^\top J X - \lambda J = 0.5\left[\begin{smallmatrix} 2ac & ad+bc-\lambda \\ ad+bc - \lambda & 2bd \end{smallmatrix}\right].
\end{align*}
By Sylvester's criterion, $X^\top J X - \lambda J$ is positive semi-definite if and only if all of its principal minors are non-negative, i.e. $ac \ge 0$, $bd \ge 0$ and $\det(X^\top J X - \lambda J) \ge 0$. From the first two inequalities we conclude that $a,c$ possess the same sign, and the same holds for $b,d$. By switching between $X,-X$, we may assume without loss of generality that $a,c$ are non-negative. If $b,d$ are also non-negative, the proof is complete. Thus, it's enough to show that if $b,d$ are non-positive (and not both zero), then for any $\lambda > 0$, $4\det(X^\top J X - \lambda J) < 0$. 
By definition, we have,
\begin{align*}
&4\det(X^\top J X - \lambda J) = -(ad+bc-\lambda)^2 +4abcd.
\end{align*}
Moreover, if $b,d$ are non-positive then $abcd \le 0$. If $abcd < 0$, then $4\det(X^\top J X - \lambda J)$ must be negative. Otherwise, the determinant can be non-negative only at $\lambda = ad+bc$, but because $a,c\ge 0$, if $b,d \le 0$ then $ad+bc$ is non-positive. In particular, if $b,d\le 0$ then the determinant is negative for all $\lambda > 0$. Thus, $\det(X^\top J X - \lambda J) \ge 0$ is equivalent to $a,b,c,d \ge 0$. This concludes the proof.
\demo
\end{pf}

\section{Applications}\label{sec.AppExam}
In this section, we consider possible applications of the achieved characterization for synthesis. We explore three possible applications, including multi-purpose transformations, passivation with respect to multiple equilibria, and optimal passifying transformations.

\subsection{Multiple Purpose Transformations} \label{subsec.MPT}
Suppose we are provided with different systems $\{\Sigma_i\}_{i\in I}$ which are I/O $(\rho_i,\nu_i)$-passive. Our goal is to design a transformation $T$ mapping each system $\Sigma_i$ to an I/O $(\rho_i^\star,\nu_i^\star)$-passive system (for $i\in I$). This problem arises in two real-world occasions. First, suppose we want to design a controller which stabilizes a certain ``plug-and-play" system, i.e. the system at hand is unknown until after the controller is designed and connected. Suppose we know, however, that the system belongs to some set $\Omega$. One can compute dissipation inequalities for all possible plants in $\Omega$, find a transformation which passivizes all of them, and then implement a strictly passive feedback controller, thus stabilizing the system. Second, suppose we wish to control a system which can fault. We wish to find a transformation which makes the faultless system as strictly passive as possible, but also passivizes any faulty version of the system. When connecting the system with a strictly passive feedback controller, the first part improves the convergence rate, and the second part ensures stability of the closed-loop system.

\begin{proposition}
Consider the MIMO systems $\{\Sigma\}_{i\in  I}$ with input- and output-dimension equal to $d$, which are I/O $(\rho_i,\nu_i)$-passive. Let $(\rho^\star_i,\nu_i^\star)$ be real numbers such that for each $i$, $\rho_i^\star\nu^\star_i < 1/4$. Consider a general I/O transformation $T$ of the form \eqref{eq.Transformation}. The transformed systems $\{\tilde{\Sigma}\}_{i\in I}$ are I/O $(\rho_i^\star,\nu_i^\star)$-passive for all $i$, if and only if there exists matrices $T_i \in GL_{2d}(\R)$, and numbers $\lambda_i \ge 0$ such that the following set of constraints holds:
\begin{equation} \label{eq.Demand1}
\begin{cases}
T = (S_{\rho_i^\star,\nu_i^\star} \otimes {\rm Id}_d)T_i (S_{\rho_i,\nu_i}^{-1} \otimes {\rm Id}_d) \\
T_i^\top J T_i - \lambda_i J \ge 0. 
\end{cases}
\end{equation}
\end{proposition}
The proposition immediately follows from Proposition \ref{thm.MIMOCharacter}. In particular, if we only wish to passivize the systems $\{\Sigma_i\}$ (i.e. $\rho_i^\star = \nu_i^\star = 0$), we get the following set of equations and inequalities:
\begin{equation} \label{eq.Demand2}
\begin{cases}
T = T_i (S_{\rho_i,\nu_i}^{-1} \otimes {\rm Id}_d) \\
T_i^\top J T_i - \lambda_i J \ge 0 .
\end{cases}
\end{equation}

\begin{remark}
If the set $I$ is finite, or that the set $\{(\rho_i,\nu_i,\rho_i^\star,\nu_i^\star)\}_{i\in I}$ contains only finitely many distinct elements, one can find a solution to \eqref{eq.Demand1} by stating an optimization problem with an arbitrary cost function and constraints of the form \eqref{eq.Demand1}. If the cost function is chosen as a coercive function, one can show that the problem has a solution whenever it is feasible, as the set of constraints is closed. See Section \ref{subsec.opt} for more on optimization problems and passivizing transformations.
\end{remark}

\begin{example}
Consider the SISO system $\Sigma$ which is the parallel interconnection of two linear and time-invariant SISO systems $\Sigma_1,\Sigma_2$ given by the transfer functions \textcolor{black}{$G_1(s) = \frac{s-1}{s+1}$ and $G_2(s) = \frac{-s^3+6s+5}{s^3+4s^2+5s+2}$}. The system $\Sigma$ is linear and time-invariant, and its transfer function can be computed to be  $G(s) = \frac{2s+3}{s^2+3s+2} = \frac{1}{s+2}+\frac{1}{s+1}$. It is easy to verify that $\Sigma$ is passive, and actually output-strictly passive with a parameter $\rho =\frac{2}{3}$. However, the component corresponding to $\Sigma_2$ is unreliable, and may fault. When it does, the transfer function changes to $G_1(s)$, which is not passive as it is non-minimum phase. However, it does have a finite $\mathcal{L}_2$-gain equal to $\max_{{\rm Re}(s) > 0} \|G_1(s)\| = 1$. It is shown in \cite{Sharf2019a} that a system with a finite $\mathcal{L}_2$-gain equal to $\beta$ is input $\nu$-passive for $\nu = -\beta^2-0.25$. Thus, the faultless system is output $\frac{2}{3}$-passive, and the faulty system is input $(-1.25)$-passive. 

Suppose we want to find a transformation $T$ that maps the faultless system to an output $2$-passive system, and the faulty system to a passive system. By \eqref{eq.Demand1}, if we define $T_1 = S_{2,0}^{-1}TS_{\frac{2}{3},0}$ and $T_2 = S_{0,0}^{-1}TS_{0,-\frac{5}{4}}$, then we want the entries of both $T_1$ and $T_2$ to have the same sign. \textcolor{black}{We choose $$ T = \begin{bmatrix} 1 & 0.4 \\ 0.4 & 0.2 \end{bmatrix}.$$ A simple computation shows that:
\begin{align*}
T_1 &= S_{2,0}^{-1}TS_{\frac{2}{3},0} = \frac{1}{15}\begin{bmatrix} 3 & 2 \\ 6 & 7 \end{bmatrix} \\
T_2 &= S_{0,0}^{-1}TS_{0,-\frac{5}{4}} = \begin{bmatrix} 0.4 & 0.4 \\ 0.12 & 0.2 \end{bmatrix},
\end{align*}
meaning both $T_1,T_2$ have entries which have the same sign. Thus, the map $T$ satisfies the requirements we established, which is a fact we now verify independently of the computation above. }

\textcolor{black}{
First, we note that the map $T$ can be written as the product $$T=\begin{bmatrix} 1 & 0 \\ 0.4 & 1 \end{bmatrix}\begin{bmatrix} 1 & 0 \\ 0 & 0.04 \end{bmatrix}\begin{bmatrix} 1 & 0.4 \\ 0 & 1 \end{bmatrix},$$ meaning it operates in the following way:  It first implements a constant output feedback with gain equal to $0.4$, it then implements a gain on the output of size $0.04$, and at the end it implements a constant input feed-through with gain $0.4$. Thus, it transforms the faultless transfer function $G$ to $$\tilde G(s) = \frac{0.04}{\frac{1}{G}+0.4}+0.4 = \frac{0.4s^2+1.6s+1.4}{s^2+3.8s+3.2},$$ and simultaneously transforms the fault transfer function $G_1(s)$ to $$\tilde G_1 (s) = \frac{0.04}{\frac{1}{G_1}+0.4}+0.4 = \frac{0.6s+0.2}{1.4s+0.6}.$$ One could easily check (e.g., using the MATLAB command ``getPassiveIndex") that $\tilde G(s)$ is output-strictly passive with index $\rho \approx 2.2857 > 2$, and that $\tilde G_1(s)$ is passive, and actually output-strictly passive with index $\rho \approx 2.333$. Thus, the transformation $T$ maps the faultless system to an output $2$-passive system, and the faulty system to a passive system, as required.}
\end{example}

\subsection{Passivation with Respect to Multiple Equilibria and Equilibrium-independent Passivity}
In several occasions, one wishes to study the behavior of a system around more than one equilibrium. One example includes consumer products that have multiple operation settings e.g., a food processor with low, middle, and high settings, or a refrigerator with multiple cooling levels possible. Another example consists of systems which need to operate under a wide range of inputs, e.g. egg-sorting machines which need to lift eggs of different sizes without breaking them, or warehouse robots which need to manipulate goods or crates of different shapes and sizes without breaking them or dropping them. A third example includes multi-agent networks, for which the steady-state limit can be hard \textcolor{black}{or even} impossible to guess before running the network due to the agents having different models, goals and restrictions. 

In this direction, one can consider passivity (or shortage thereof) with respect to an arbitrary steady-state I/O pair $(\mathrm u,\mathrm y)$. Indeed, the notions of output $\rho$-passivity, input $\nu$-passivity and I/O $(\rho,\nu)$-passivity can be extended to other steady-states by replacing $y(t)$ by $y(t)-\mathrm y$ and $u(t)$ by $u(t) - \mathrm u$ in \eqref{eq.OWPassivity}, \eqref{eq.IWPassivity}, and \eqref{eq.IOWPassivity} respectively. When designing controllers for systems which can operate around more than one equilibrium, we need to consider passivity (or I/O $(\rho,\nu)$-passivity) with respect to each equilibrium. The same system can behave differently around different equilibria. e.g. the static nonlinearity $y = \frac{u}{u^2+1}$ is passive around the steady-state I/O pair $(0,0)$, but is not passive around the steady-state I/O pair $(1,0.5)$. To remedy this problem, the notions of equilibrium-independent passivity [\cite{Hines2011}] and maximal equilibrium-independent passivity [\cite{Burger2014,Sharf2018a}] were offered. Under these assumptions, it is possible to prove that \textcolor{black}{certain complex systems (e.g., multi-agent networks)} converge without specifying a limit ahead of time. In some cases, we might know that there are some $\rho,\nu$ such that the system is I/O $(\rho,\nu)$-passive with respect to all equilibria, and in that case we can use the method of \cite{Sharf2019a} to passivize the system with respect to all equilibria. However, we can consider a more general case, where different equilibria are associated with different corresponding dissipation inequalities. 

Before moving forward, we note that the inequalities defining I/O $(\rho,\nu$)-passivity with respect to any steady-state can also be written as ($d$-dimensional) PQIs in exactly the same way used for I/O $(\rho,\nu)$-passivity with respect to the origin. Thus, our results characterize all transformations $T$ that passivize a plant with respect to any fixed equilibrium. In that direction, we consider a system $\Sigma$ and a collection of steady-state I/O pairs $\{(\mathrm u_i,\mathrm y_i)\}$. Our goal is to find a transformation $T$ that passivizes the system $\Sigma$ with respect to all (transformed) steady-state pairs simultaneously.

\begin{remark}
Intuitively, one might try to keep the same steady-state I/O pairs for the transformed system. However, this might be impossible if the transformed system must be passive. For example, if we have a SISO system $\Sigma$ with two steady-state I/O pairs $(0,0)$ and $(1,-1)$, then the system cannot be passive, as the corresponding steady-state relation is non-monotone.
\end{remark}

Unsurprisingly, this problem is very similar to the multiple objective transformation considered in Section \ref{subsec.MPT}. We can prove the following proposition:

\begin{proposition}
Consider a MIMO system $\Sigma$ with input- and output-dimension equal to $d$. Let $\{(\mathrm u_i,\mathrm y_i)\}$ be a collection of steady-state I/O pairs of $\Sigma$, and let $(\rho_i,\nu_i),(\rho^\star_i,\nu_i^\star)$ be real numbers such that for each $i$, $\rho_i\nu_i, \rho_i^\star\nu^\star_i < 1/4$. Suppose that for each $i$, the system $\Sigma$ is I/O $(\rho_i,\nu_i)$-passive with respect to the steady-state I/O pair $(\mathrm u_i,\mathrm y_i)$. Consider a general I/O transformation $T$ of the form \eqref{eq.Transformation}, and consider the new system $\tilde{\Sigma}$ and the new steady-state pairs $\{T(\mathrm u_i,\mathrm y_i)\}$. $\tilde{\Sigma}$ is I/O $(\rho_i^\star,\nu_i^\star)$-passive with respect to $T(\mathrm u_i,\mathrm y_i)$, for all $i$, if and only if there exists matrices $T_i \in GL_{2d}(\R)$, and numbers $\lambda_i \ge 0$ such that the following set of constraints holds:
\begin{equation} \label{eq.Demand1a}
\begin{cases}
T = (S_{\rho_i^\star,\nu_i^\star} \otimes {\rm Id}_d)T_i (S_{\rho_i,\nu_i}^{-1} \otimes {\rm Id}_d) \\
T_i^\top J T_i - \lambda_i J \ge 0 .
\end{cases}
\end{equation}
\end{proposition}
As before, the proof of the proposition follows immediately from Proposition \ref{thm.MIMOCharacter}. We again note that when we wish to passivize the system $\Sigma$ with respect to all equilibria (i.e., $\rho_i^\star,\nu_i^\star = 0$), we get the following set of equations and inequalities:
\begin{equation} \label{eq.Demand2a}
\begin{cases}
T = T_i (S_{\rho_i,\nu_i}^{-1} \otimes {\rm Id}_d) \\
T_i^\top J T_i - \lambda_i J \ge 0. \nonumber
\end{cases}
\end{equation}
\subsection{Optimal Passivizing Transformations} \label{subsec.opt}
In the previous sections of the paper, we characterized all transformations that passivize a given system $\Sigma$. Thus, it is natural to ask questions such as ``which passivizing transformation minimizes (or maximizes) a given quantity?" One class of quantities of interest can be system-theoretic properties of the transformed system, e.g. the $\mathcal{L}_2$-gain or tracking error for a given input and a desired output. Another class of interesting quantities to optimize consists of properties of $T$. These include, for example, the distance of the transformation $T$ from a nominal transformation $T_0$, e.g. the identity. One could also consider ``mixed" quantities, e.g. the distance in the $\mathcal{H}_\infty$-norm between the original system and the transformed system.

Generally, one could consider a transformation $T$ that maps a given I/O $(\rho,\nu)$-system to an I/O $(\rho_\star,\nu_\star)$ system. The quantity we wish to minimize can be written as a function $\Phi(T)$ of $T$. The associated optimization problem reads:
\begin{align*}
\min_T \quad & \Phi(T) \\
{\rm s.t.}\quad & \text{$T$ maps I/O $(\rho,\nu)$ systems} \\
\quad & \text{to I/O $(\rho_\star,\nu_\star)$-systems.}
\end{align*}
One could use Theorem \ref{thm.MIMOCharacter} to restate the optimization problem in a tractable form:
\begin{align} \label{eq.NontractableMIMO}
\min_{T,\lambda,M} \quad & \Phi(T) \\ \nonumber
{\rm s.t.}\quad & M = (S_{\rho^\star,\nu^\star} \otimes {\rm Id}_d)^{-1} T (S_{\rho,\nu} \otimes {\rm Id}_d) \\ \nonumber
\quad & M^\top J M - \lambda J \ge 0 \\ \nonumber
\quad & \lambda \ge 0,
\end{align}
where $J = \left[\begin{smallmatrix} 0 & 1/2 \\ 1/2 & 0 \end{smallmatrix}\right]$. This optimization problem can be easily defined for any cost function $\Phi$, whether it is explicitly defined using a formula involving $T$, or implicitly defined by a characteristic of the transformed system $\tilde{\Sigma}$. However, solving the optimization problem can be hard. First, the function $\Phi$ might not be explicitly given, or non-convex. Second, even if the function $\Phi$ was convex, the constraint $M^\top J M - \lambda J \ge 0$ is non-convex, as the matrix $J$ is not positive semi-definite.
 We should note, however, that the latter problem can be easily remedied for SISO systems. Indeed, by Proposition \ref{prop.SameSign}, the constraint can be replaced by:
\begin{align*}
\mbox{i) }&T = \left[\begin{smallmatrix} a & b \\ c & d \end{smallmatrix}\right] \\
\mbox{ii) }&a,b,c,d \text{ have the same sign}.
\end{align*}
This constraint is still non-convex, but can be convexified by separating the problem into two sub-problems, one with the constraint $a,b,c,d \ge 0$, and one with the constraint $a,b,c,d \le 0$. 
\begin{remark}
Returning to the MIMO case, one can prove that the matrix $M = \left[\begin{smallmatrix} \alpha{\rm Id}_d & \beta{\rm Id}_d \\ \gamma{\rm Id}_d & \delta{\rm Id}_d\end{smallmatrix}\right] \in \mathbb{R}^{2d\times 2d}$ satisfies the inequality $M^\top J M - \lambda J \ge 0$ for some $\lambda > 0$ whenever $\alpha,\beta,\gamma,\delta$ have the same sign, similarly to Proposition \ref{prop.SameSign}. Thus, one can consider a tractable relaxation of the optimization problem \eqref{eq.NontractableMIMO} by similarly replacing the constraints $M^\top J M - \lambda J \ge 0, \lambda \ge 0$ by the constraint $M = \left[\begin{smallmatrix} \alpha{\rm Id}_d & \beta{\rm Id}_d \\ \gamma{\rm Id}_d & \delta{\rm Id}_d\end{smallmatrix}\right] \in \mathbb{R}^{2d\times 2d}$ and demanding that $\alpha,\beta,\gamma,\delta$ have the same sign.
\end{remark}

We now give examples of two tractable optimization problems:

\begin{example}
Consider the problem of transforming an SISO I/O $(\rho,\nu)$-passive system to an I/O $(\rho_\star,\nu_\star)$-passive system. We wish to find such a transformation which is closest to a given transformation $T_0$, i.e., minimizes the operator norm $\|T-T_0\|$. By the discussion above, we can write the problem as:
\begin{align*}
\min_{T,M,\lambda} \quad & \|T-T_0\| \\
{\rm s.t.}\quad & M = (S_{\rho^\star,\nu^\star} \otimes {\rm Id}_d)^{-1} T (S_{\rho,\nu} \otimes {\rm Id}_d) \\
\quad & M^\top J M - \lambda J \ge 0 \\
\quad & \lambda \ge 0
\end{align*}
However, minimizing $\|T-T_0\|$ directly is hard. Instead, we introduce a new variable $\gamma$  and demand that $(T-T_0)(T-T_0)^\top \le \gamma {\rm Id}_{2}$, so that minimizing $\gamma$ gives the desired result (and the operator norm $\|T-T_0\|$ is given by $\sqrt{\gamma}$. One can rewrite the last inequality as a linear matrix inequality using Schur's complements, giving the following equivalent optimization problem:
\begin{align} \label{eq.GeneralOptimizationClosest}
\min_{T,M,\lambda,\gamma} \quad & \gamma \\ \nonumber
{\rm s.t.}\quad & M = (S_{\rho^\star,\nu^\star} \otimes {\rm Id}_d)^{-1} T (S_{\rho,\nu} \otimes {\rm Id}_d) \\ \nonumber
\quad & M^\top J M - \lambda J \ge 0 \\ \nonumber
\quad & \begin{bmatrix} {\rm Id}_2 & T-T_0 \\ T^\top - T_0^\top & \gamma {\rm Id}_2\end{bmatrix} \ge 0\\ \nonumber
\quad & \lambda \ge 0. \nonumber
\end{align}
where the matrix inequalities are understood as LMIs (rather than elementwise inequalities).

As a concrete example, we take $(\rho,\nu) = (0,-1)$ and $(\rho_\star,\nu_\star) = (1,0)$. We thus seek a transformation of the form \eqref{eq.Transformation} mapping an input passive-short SISO system with parameter $\nu = -1$ to an output strictly-passive system with $\rho = 1$. Classically, one would first use feed-through to passivize the system, and then implement a feedback to increase its output passivity. This results in the transformation 
$
T = \left[\begin{smallmatrix} 1 & 0 \\ 1 & 1 \end{smallmatrix}\right]\left[\begin{smallmatrix} 1 & 1 \\ 0 & 1 \end{smallmatrix}\right] = \left[\begin{smallmatrix} 1 & 1 \\ 1 & 2 \end{smallmatrix}\right].
$
We wish to find such a transformation which is closest to the identity transformation, i.e., minimizes the operator norm $\|T-{\rm Id}_2\|$. Using Proposition \ref{prop.SameSign}, \eqref{eq.GeneralOptimizationClosest} is recast as:
\begin{align*}
\min_{T,\gamma,a,b,c,d} \quad & \gamma \\
{\rm s.t.}\quad & a,b,c,d\text{ have the same sign} \\
\quad& T = \begin{bmatrix} 1 & 1 \\ 0 & 1 \end{bmatrix}\begin{bmatrix} a & b \\ c & d \end{bmatrix}\begin{bmatrix} 1 & 0 \\ 1 & 1 \end{bmatrix},\\ \quad&
\begin{bmatrix} {\rm Id}_2 & T-{\rm Id}_2 \\ T^\top - {\rm Id}_2 & \gamma {\rm Id}_2\end{bmatrix} \ge 0.
\end{align*}
This problem is non-convex, but it can be written as the minimum of cone programming problems, one with the constraint $a,b,c,d\ge 0$, and one with the constraints $a,b,c,d \le 0$. We can thus solve the problem by computer and find the optimal transformation $T = \left[\begin{smallmatrix} 1.5 & 0.5 \\ 0.5 & 0.5\end{smallmatrix}\right]$, which corresponds to $a=1, b=c=0, d = 0.5$. The operator norm in this case is $\sqrt{\gamma} = \frac{1}{\sqrt{2}}\approx 0.707$. As a comparison, the operator norm $\|T-I\|$ for $T = \left[\begin{smallmatrix} 1 & 1 \\ 1 & 2 \end{smallmatrix}\right]$ is $\frac{\sqrt{3+\sqrt{5}}}{\sqrt{2}}\approx 1.618$.
\end{example}

\begin{figure}[b]
    \centering
    \includegraphics[scale = 0.8]{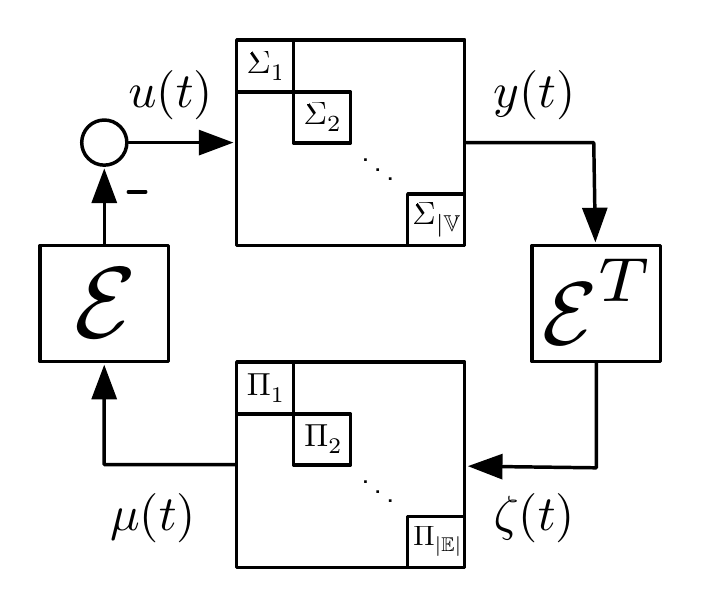}
    \caption{Diffusively coupled network.}
    \label{fig.nds}
\end{figure}

\begin{example}
\textcolor{black}{
Consider the problem of transforming a given SISO LTI system $\Sigma$ with transfer function $G(s)$ to a passive LTI system. We know that the system is I/O $(\rho,\nu)$-passive, where we assume $\rho,\nu \le 0$ and aim to minimize the $\mathcal{H}_\infty$-norm of the transformed system. 
It is clear that the $\mathcal{H}_\infty$ norm of the transformed system can be forced to be arbitrarily small by applying a pre-gain or post-gain, without harming the passivity of the transformed system. Thus, to make the problem non-degenerate, we focus on transformations of a specific form, namely, transformations $T$ that can be achieved using a single feedback block, and a single feed-through block. These transformations can be seen in Fig \ref{fig.BlockDiagram}, where the pre-gain $\delta_D$ and the post-gain $\delta_B$ are taken to be equal to $1$. Thus, the transformation $T$ can be described as:
\begin{align*}
    T = \begin{bmatrix}
    1 & \delta_C \\ \delta_A & 1+\delta_A\delta_C
    \end{bmatrix}
\end{align*}
where $\delta_A$ is the feedback gain and $\delta_C$ is the feed-through gain, as in Fig. \ref{fig.BlockDiagram}}

\textcolor{black}{
Writing $T = \left[\begin{smallmatrix}1 & b \\ c& 1+bc \end{smallmatrix}\right]$ for numbers $b,c\in \R$, the auxiliary input is given by $\tilde{u} = u+by$ and the auxiliary output is $\tilde{y} = cu+(1+bc)y$. Thus, the transfer function of the transformed system is $\tilde{G}(s) = \frac{c+(1+bc)G(s)}{1+bG(s)}$. Using Theorem \ref{theorem.SISOProduct}, the optimization problem has the form:
\begin{subequations}\label{eq.16}
\begin{align} \label{eq.16a}
\min_{b,c,\theta,M} \quad & \max_{\omega \in \R} \left\|\frac{c+(1+bc)G(j\omega)}{1+bG(j\omega)}\right\| \\ \label{eq.16b}
{\rm s.t.} \quad& \left[\begin{smallmatrix}
1 & b \\ c & 1+bc
\end{smallmatrix}\right] = \theta M S_{\rho,\nu}^{-1}\\
~&~ M_{11},M_{12},M_{21},M_{22} \ge 0 ,~~\theta \in \{\pm 1\} .\nonumber
\end{align}
\end{subequations}}
\textcolor{black}{
We can derive explicit constraints on $b,c$ from \eqref{eq.16b}. We focus on the case $\nu<0$, whereas the case $\nu = 0$ can be derived similarly. In this case, we have
\begin{align*}
S_{\rho,\nu} = \begin{bmatrix} \frac{-1-R}{2\nu} & \frac{1-R}{2\nu} \\ -1 & 1 \end{bmatrix},~
    S_{\rho,\nu}^{-1} = \begin{bmatrix} -\nu & \frac{1-R}{2} \\ -\nu & \frac{1+R}{2} \end{bmatrix},
\end{align*}
where $R = \sqrt{1-4\rho\nu} \in [0,1]$. The equation \eqref{eq.16b} implies that $1=-\theta\nu(M_{11}+M_{12})$, i.e., that $\theta = 1$ as $\nu < 0$. We can write \eqref{eq.16b} as 
\begin{align*}
    M &= \begin{bmatrix} 1 & b \\ c & 1+bc \end{bmatrix}\\
    S_{\rho,\nu} &= \begin{bmatrix} \frac{1+R}{-2\nu} - b & \frac{1-R}{2\nu}+b \\ \frac{1+R}{-2\nu}c - (1+bc) & \frac{1-R}{2\nu}c + 1 +bc\end{bmatrix},
\end{align*}
which allows us to rewrite \eqref{eq.16b} as four inequalities in the variables $b,c$. Namely, \eqref{eq.16} is recast as:
\begin{align*}
\min_{b,c} \quad & \max_{\omega \in \R} \left\|\frac{c+(1+bc)G(j\omega)}{1+bG(j\omega)}\right\| \\ 
{\rm s.t.} \quad& \frac{1-R}{-2\nu} \le b \le \frac{1+R}{-2\nu},~ c\ge \frac{1}{\frac{1+R}{-2\nu}-b}.
\end{align*}
Observe that there are only three inequalities here as the fourth one is written as $c \ge -(b-\frac{1-R}{-2\nu})^{-1}$, which is redundant given the other constraints. }

\textcolor{black}{Moreover, we claim that as $c$ grows, the cost function grows larger. Indeed, if $c_1 < c_2$ and we let $y_1 = y + c_1u$, $y_2 = y+c_2u = y_1+(c_2-c_1)u$ be the associated outputs, and we assume that the system with $c_1$ is already passive, then for any input $u$ we have:
\begin{align*}
    \langle y_2,y_2 \rangle &= \langle y_1,y_1 \rangle + 2(c_2-c_1) \langle u,y_1 \rangle + (c_2-c_1)^2 \langle u,u\rangle,\\
    &\ge \langle y_1,y_1 \rangle + (c_2-c_1)^2 \langle u,u\rangle,
\end{align*}
as $\langle u,y_1\rangle \ge 0$ from passivity, where $\langle \cdot,\cdot\rangle$ is the inner product of $\mathcal{L}_2$ signals. Thus, we can restrict the value of $c$ the lower bound $(\frac{1+R}{-2\nu}-b)^{-1}$. Some algebra shows that the optimization problem is then recast as:
\begin{align} \label{eq.18}
\min_{b} \quad & \left[\frac{-2\nu}{1+R+2\nu b}~\max_{\omega \in \R} \left\|\frac{1+\frac{1+R}{-2\nu}G(j\omega)}{1+bG(j\omega)}\right\|\right] \\ \nonumber
{\rm s.t.} \quad &~ \frac{1-R}{-2\nu} \le b \le \frac{1+R}{-2\nu}.
\end{align}
This is now a (nonlinear) optimization problem in one variable, $b$, which is contained in a bounded interval. Thus, the problem can be solved efficiently by computing the cost function on a tight grid of values of $b$.}

\textcolor{black}{
As a specific example, we take $G(s) = \frac{\kappa s-1}{s+1}$ for some parameter $\kappa > 0$, which is I/O $(0,-1)$-passive, and show the problem can be solved analytically. Here, $\rho = 0, \nu = -1$ and $R = 1$, implying that $\frac{1-R}{-2\nu} = 0$ and $\frac{1+R}{-2\nu} = 1$. In particular, the associated system is not passive as $G(s)$ is not minimum-phase. Some algebra shows that the transformed transfer function is given by:
\begin{align*}
    \frac{1+\frac{1+R}{-2\nu}G(j\omega)}{1+bG(j\omega)} = \frac{1+G(j\omega)}{1+bG(j\omega)} = \frac{(1+\kappa)j\omega}{(1+\kappa b)j\omega + (1-b)}.
\end{align*}
Thus, the cost function of \eqref{eq.18} is given by:
\begin{align*}
    \frac{1}{1-b} \cdot \max_{\omega \in \R} \frac{(1+\kappa)\omega}{\sqrt{(1+\kappa b)^2 \omega^2 + (1-b)^2}}.
\end{align*}
By dividing the numerator and the denominator of the quotient by $\omega^2$, it is clear that the maximum is achieved either for $\omega = 0$ or for $\omega = \infty$. Thus, \eqref{eq.18} is given by:
\begin{align*}
    \min_{0\le b \le 1}~\max \left\{\frac{1}{(1-b)^2},\frac{1+\kappa}{(1-b)(1+\kappa b)}\right\}
\end{align*}
The first element is the larger one if $b \ge \frac{\kappa}{2\kappa+1}$, and the second element is larger if $b \le \frac{\kappa}{2\kappa + 1}$. For that reason, we solve the problems:
\begin{align*}
    \min_{0\le b \le \frac{\kappa}{2\kappa+1}} \frac{1+\kappa}{(1-b)(1+\kappa b)},~
    \min_{\frac{\kappa}{2\kappa+1}\le b \le 1} \frac{1}{(1-b)^2}
\end{align*}
The second problem is always minimized at $b = \frac{\kappa}{2\kappa +1}$. The first problem can be understood as maximizing the quadratic function $\kappa (1-b)(\kappa^{-1}+ b)$, which is maximized globally at $b = \frac{1-\kappa^{-1}}{2}$. This point is always smaller than the upper bound $\frac{\kappa}{2\kappa+1}$, but might drop below the lower bound $0$ if $\kappa < 1$. }

\textcolor{black}{
To conclude, we must check two options for the optimal value of $b$. If $\kappa < 1$, the options are $b=0$ (with a cost of ${1+\kappa}$) and $b=\frac{\kappa}{2\kappa+1}$ (with a cost of $\frac{(2\kappa+1)^2}{(\kappa+1)^2}$, which is higher if $\kappa < 1$). Thus, the optimal transformation is achieved for $b=0$ and $c=\frac{1}{\frac{1+R}{-2\nu}-b} = 1$. The associated transformed system has the transfer function $\tilde{G}(s) = G(s) + 1 = \frac{(\kappa+1)s}{s+1}$, having an $\mathcal{H}_\infty$ norm of $1+\kappa$.}

\textcolor{black}{
If instead we have $\kappa \ge 1$, the two options are $b = \frac{1-\kappa^{-1}}{2}$ (with a cost of $\frac{4\kappa}{1+\kappa}$), and and $b=\frac{\kappa}{2\kappa+1}$ (with a cost of $\frac{(2\kappa+1)^2}{(\kappa+1)^2}$), which is always higher). Thus, the optimal transformation is achieved for $b=\frac{1-\kappa^{-1}}{2}$ and $c=\frac{1}{\frac{1+R}{-2\nu}-b} = \frac{2\kappa}{\kappa+1}$, and the associated transformed system has the transfer function $\tilde{G}(s) = \frac{G(s) + 1}{bG(s)+1} = \frac{2(\kappa+1)s}{(\kappa+1)s+1+\kappa^{-1}}$, having an $\mathcal{H}_\infty$ norm of $\frac{(2\kappa+1)^2}{(\kappa+1)^2}$.}
\end{example}
\vspace{-20pt}
\textcolor{black}{
\section{Case Study: Synchronization for Faulty Non-Passive Systems} \label{sec.Synch}}
\textcolor{black}{
We consider a network of SISO agents $\{\Sigma_i\}_{i\in V}$. Our goal is to synthesize controllers which synchronize the agents, i.e., assert that they asymptotically converge to the same output. In order to do so, we adopt the formalism of diffusively-coupled networks (see, e.g., \cite{Burger2014}). In that formalism, the interconnection topology between the agents is described using a graph $\G = (V,E)$, where $V$ is the set of vertices and $E$ is the set of edges, whose elements are pairs of vertices $e=(i,j)$, where $i$ is the \emph{tail} of the edge, and $j$ is the \emph{head} of the edge. The graph gives rise to the incidence matrix $\E \in \R^{|V|\times|E|}$, defined as follows - for any edge $e=(i,j)$, we have $\E_{ie}=-1, \E_{je}=1$ and $\E_{ke}=0$ for any other vertex $k\in V$. The vertices of the graph correspond to the agents, and the edges in the graph represent (possibly dynamic) controllers $\{\Pi_e\}_{e\in E}$ that regulate the relative output between the corresponding incident agents. We denote the input and the output to the $i$-th agent by $u_i$ and $y_i$ respectively, and the input and the output to the $e$-th edge controller by $\zeta_e$ and $\mu_e$ respectively. We denote the stacked signals by $u=[u_1^\top,\ldots,u_{|V|}]^\top$, and similarly for $y,\zeta$ and $\mu$. The loop between the agents and the controllers is closed by taking the controller input $\zeta$ as the vector of relative outputs $\E^\top y$, and the agent input $u$ is chosen as minus the signed sum of the controller inputs, $u=-\E\mu$. The closed-loop system is pictured in Fig. \ref{fig.nds}. Passivity-based control of diffusively-coupled systems has been studied extensively over the last decade and a half, starting with the seminal paper by \cite{Arcak2007}. Later, variants of passivity, such as maximal equilibrium-independent passivity, have been applied to analyze diffusively-coupled systems and synthesize distributed controllers for them, see e.g. \cite{Burger2014} and \cite{Sharf2018a}.}

\textcolor{black}{For our problem, we consider identical agents $\Sigma_i$ which can fault mid-run. The nominal behaviour of the agents is described by the system $\Upsilon_0$, and the possible faulty modes of the agents are described by the systems $\Upsilon_1,\ldots,\Upsilon_M$, where $M$ is the number of possible faulty modes. We emphasize that the agents can fault independently of one another, i.e., the faults are local in nature. This modeling for the agents renders them as switched systems. For switched systems, in which we assume the swicthing cannot be controlled, the first results about passivity-based control required all modes to be passive with respect to a common storage function (\cite{Haddad2000,Chen2005}). More recent results have allowed one to consider switched systems in which different storage functions are associated with different modes, and some assumptions are made on the behaviour of the storage function of mode $i$ under the trajectories of mode $j\neq i$, see \cite{Zhao2008,Zakeri2019}. As this is not the main focus of this paper, we will assume that all modes $\Upsilon_i$ are input-output $(\rho_i,\nu_i)$-passive with the same storage function. The results described later can also be generalized to the case in which different fault modes are associated with different storage functions.}

\begin{figure*}
    \centering
    \subfigure[Output of the agents.] {\scalebox{.5}{\includegraphics{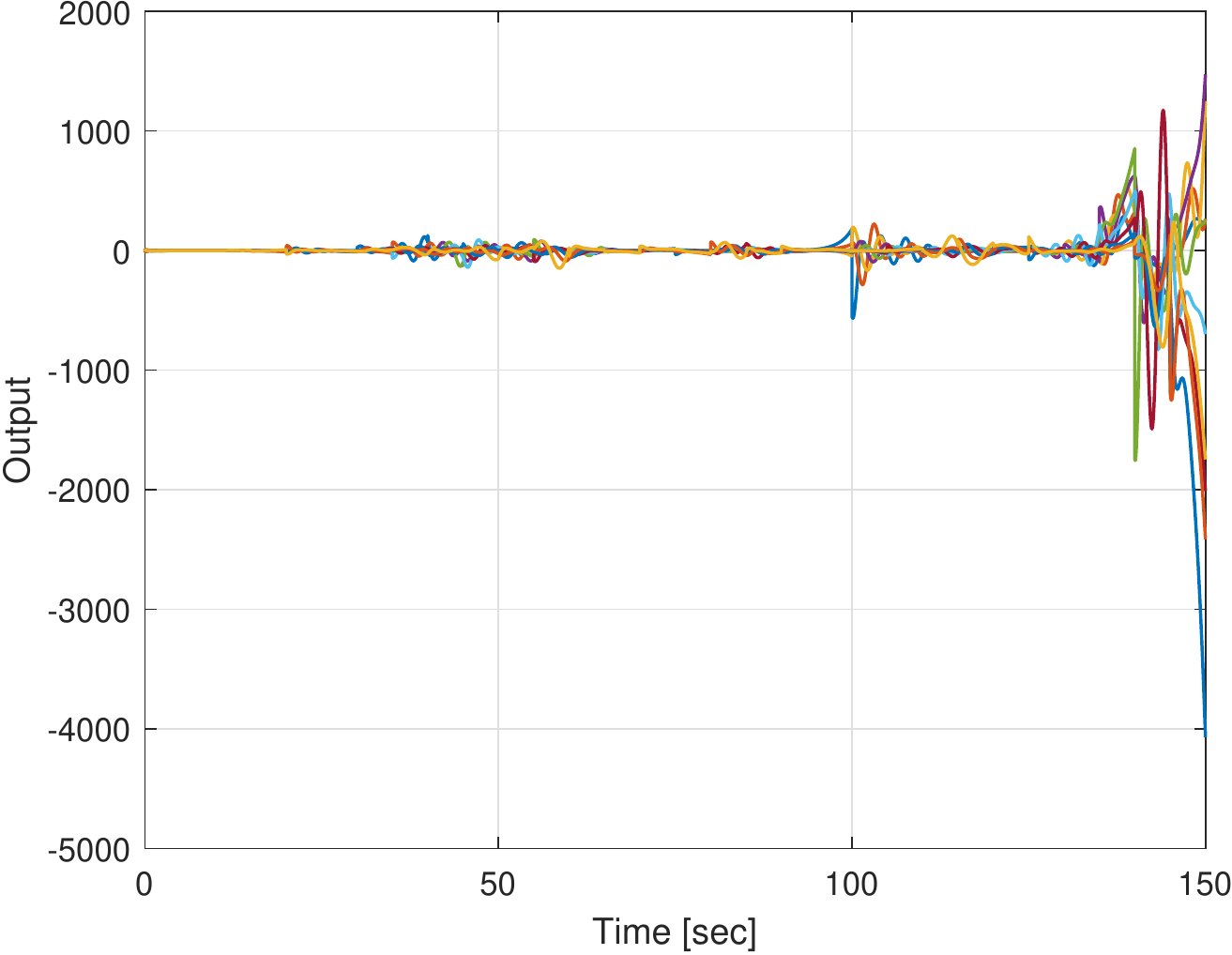}}} 
    \hfill
   \subfigure[Distance of the output $y_i(t)$ from the mean $\frac{1}{N} \sum_{i=1}^N y_i(t)$, in log-scale.] {\scalebox{.5}{\includegraphics{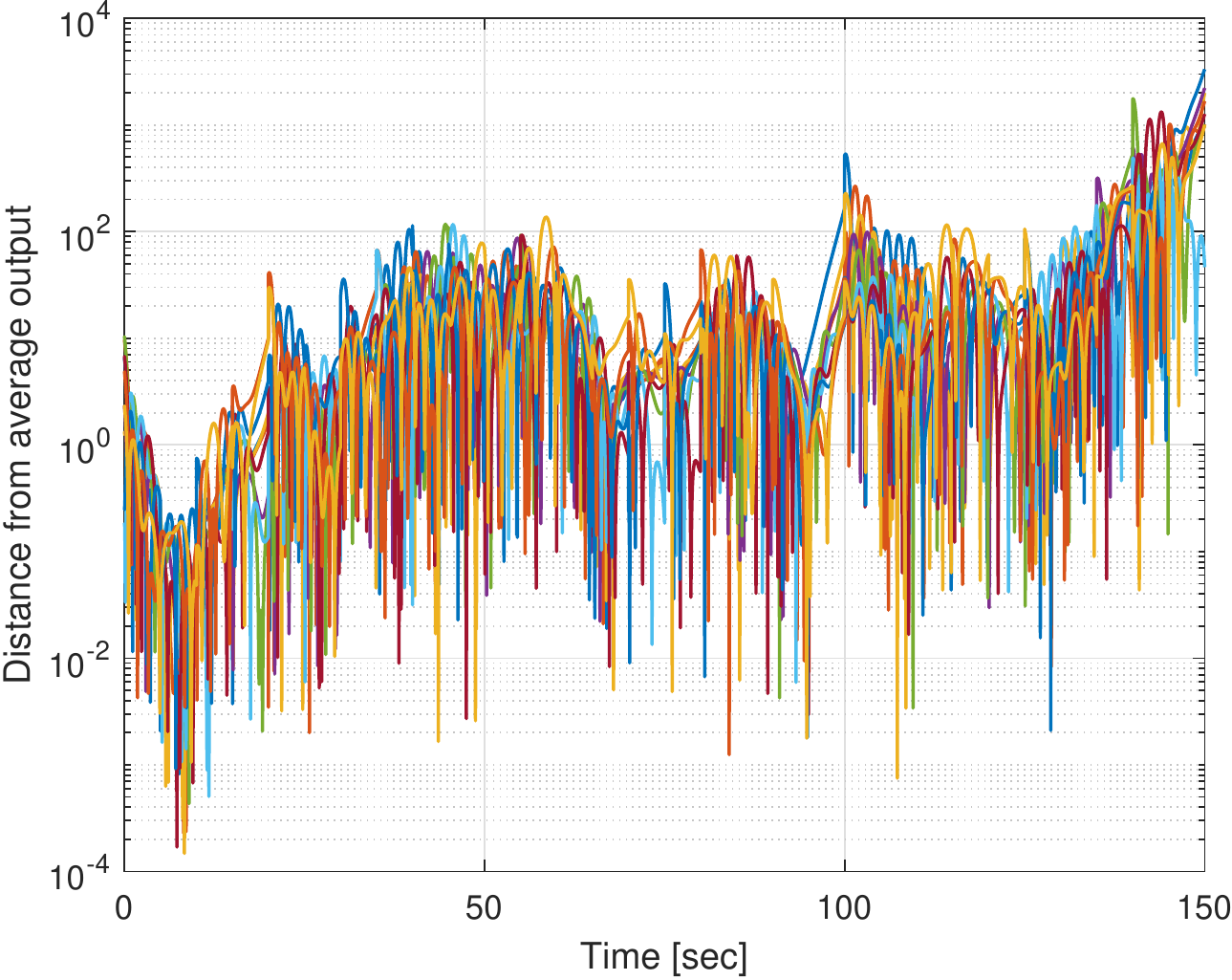}}}

    \caption{Simulation of case study in Section \ref{sec.Synch}, without passivizing the agents. It can be seen that the agents do not synchronize, and in fact stray far away from consensus in the end of the simulation.}
    \label{fig.Simulation_Nonpassivized}
\end{figure*}

\textcolor{black}{In order to synchronize these nonlinear and faulty agents, we allow the designer to implement both a local control law on individual agents, as well as networked controllers over the edges defined by the graph $\mathcal{G}$. We assume the networked controllers $\Pi_e$ are output-strictly passive with respect to the origin. Moreover, we make a certain regularity assumption on the controller, namely that if $\mu_e(t) \to 0$ then $\zeta_e(t) \to 0$. This is stronger than assuming that the origin $(0,0)$ is a steady-state input-output pair. However, this assumption holds for all LTI systems for which the control matrix $B$ has no kernel. More generally, this holds whenever we can find a (possibly nonlinear) state-space model which is observable, and the state dynamics are given in control-affine form $\dot{\xi}_e = a_e(\xi_e) + A_e(\xi_e) \zeta_e$, where the functions $a_e,A_e$ are continuous, $a_e(0) = 0$ and $A_e(0)$ has full row rank. 
}

\textcolor{black}{ 
We now prescribe a possible method to synchronize said agents using the characterization prescribed in this work. We assume the systems $\Upsilon_i$ are I/O $(\rho_i,\nu_i)$-passive for any $i=0,1,\ldots,M$, and prove the following proposition:}
\textcolor{black}{
\begin{prop} \label{prop.FDINetwork}
Suppose the assumptions above hold, and that the fault models $\Upsilon_i$ are I/O $(\rho_i,\nu_i)$-passive for any $i=0,1,\ldots,M$. Suppose that the transformation $T$  is $C_{\rho_i,\nu_i}\hookrightarrow C_{0,0}$ for any $i=0,1,\ldots,M$, so that $T$ simultaneously passivizes all fault models $\Upsilon_i$.
If we choose local control laws implementing the transformation $T$ at each agent $\Sigma_i$, then the agents synchronize, i.e. we have $\lim_{t\to \infty} (y_i(t) - y_j(t)) = 0$ for any $i,j\in V$.
\end{prop}}
\vspace{-15pt}

\begin{pf}
\textcolor{black}{
Let us denote the result of applying the transformation $T$ on $\Upsilon_0,\ldots,\Upsilon_M$ as $\tilde \Upsilon_0,\ldots,\tilde \Upsilon_M$. By assumption, all the systems $\tilde \Upsilon_0,\ldots,\tilde \Upsilon_M$ are passive with respect to a common storage function. We dnote by $\tilde{u}_i, \tilde{y}_i$ denote the input and the output of the transformed agent $\tilde{\Sigma_i}$, so that the loop is closed via $\tilde{u} = -\mathcal{E}\mu, \zeta = \mathcal{E}^\top \tilde{y}$.
By \cite[Theorem 3.4]{Burger2014} we conclude that $\mu(t) \to 0$, and the assumption on the edge controllers implies that $\zeta(t) \to 0$. Thus, the equation $\zeta(t) = \mathcal{E}^\top \tilde{y}(t)$ implies that the transformed outputs synchronize. Moreover, the equation  $\tilde u(t) = -\mathcal{E}\mu(t)$ implies that the transformed inputs also synchronize (at the value 0). By applying the inverse transformation $T^{-1}$, we conclude that the original inputs $u_i$ and the original outputs $y_i$ also synchronize. This concludes the proof. \demo}
\end{pf}
\vspace{-35pt}
\textcolor{black}{
\begin{rem}
One can similarly prove that if all of the modes $\tilde{\Upsilon}_j$ are output-strictly passive, then the output of the agents $y(t)$ converges to zero.
\end{rem}}

\begin{figure*}
    \centering
    \subfigure[Output of the agents.] {\scalebox{.5}{\includegraphics{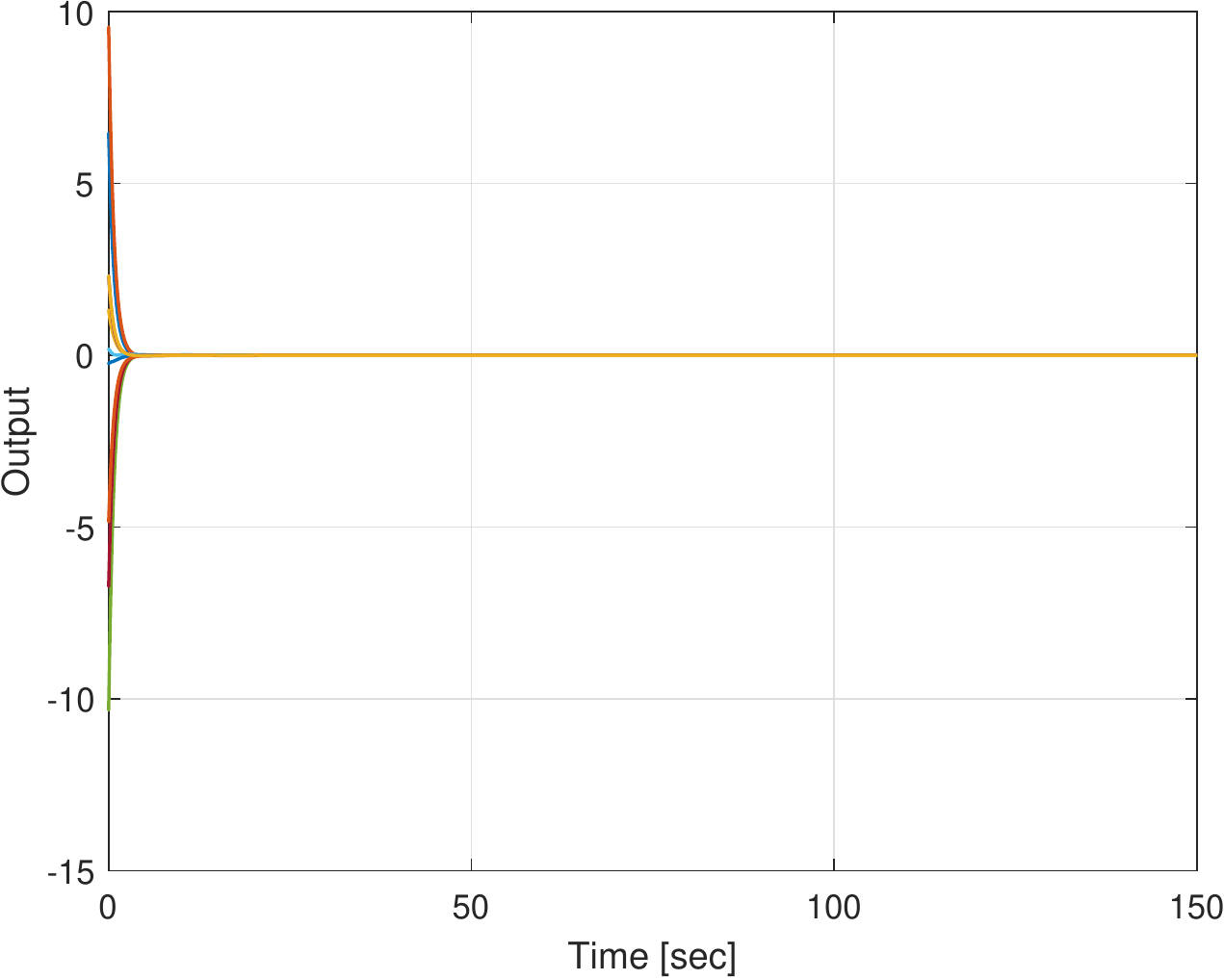}}} 
    \hfill
   \subfigure[Distance of the output $y_i(t)$ from the mean $\frac{1}{N} \sum_{i=1}^N y_i(t)$, in log-scale.] {\scalebox{.5}{\includegraphics{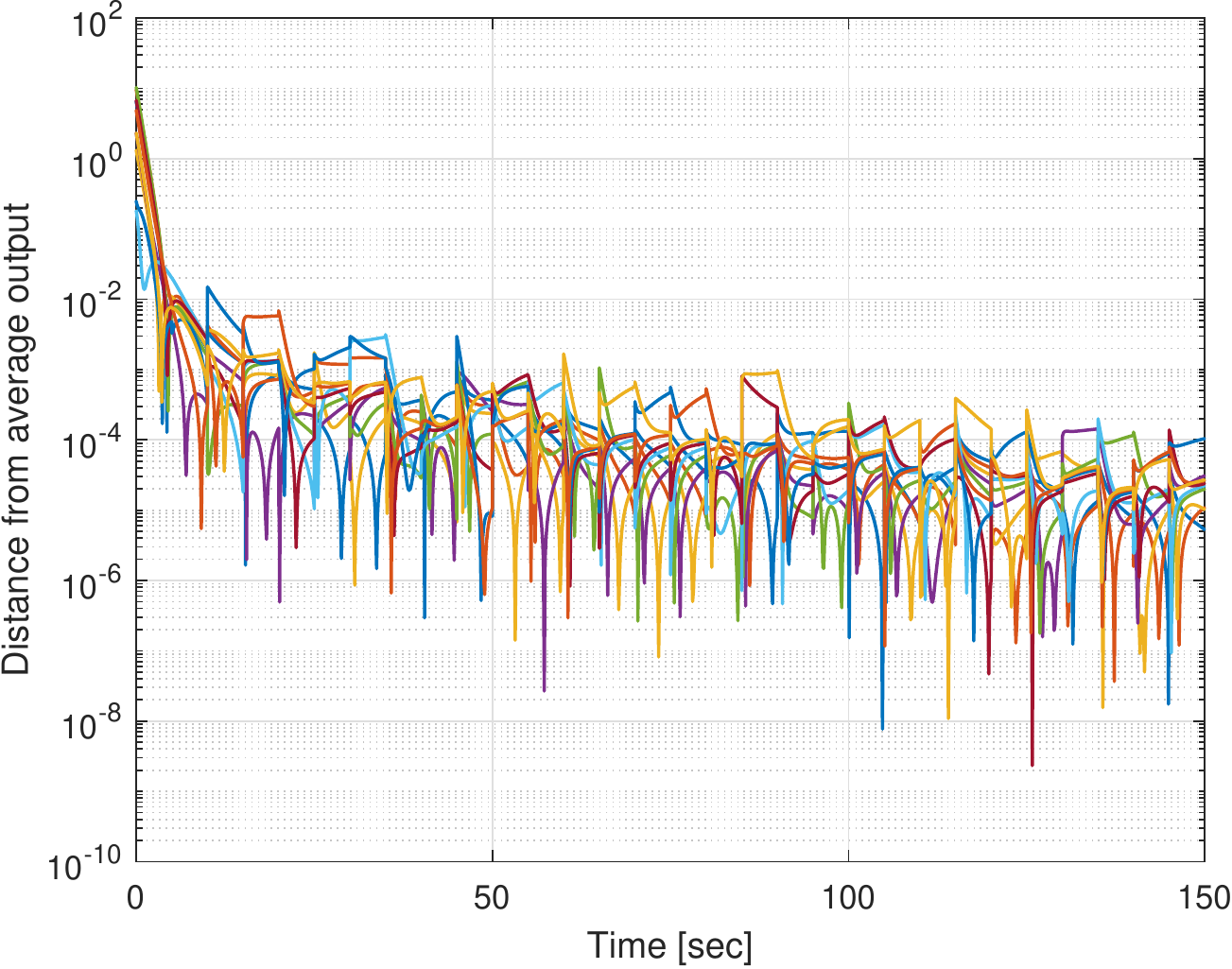}}}

    \caption{Simulation of case study in Section \ref{sec.Synch}, after applying the passivizing transformation. It can be seen that the agents synchronize.}
    \label{fig.Simulation_Passivized}
\end{figure*}

\textcolor{black}{
The proposition shows that if we choose a specific transformation $T$ which simultaneously passivizes all possible faulty modes of the agents, implement it as a local control law at every agent, and choose the edge controllers as output strictly-passive with some additional properties, then the agents synchronize. Finding the transformation $T$ itself can be done by applying the methods of the previous section.}

\textcolor{black}{
We now demonstrate the result of this theorem, as well as other results in this paper, using a simulation. We consider a network of $N=10$ agents, connected using a cycle graph. We consider agents with a nominal behaviour $\Upsilon_0$ which is LTI with transfer function $G_0(s) = \frac{2s+3}{(s+1)(s+2)}$. The faulty modes $\Upsilon_1,\Upsilon_2,\Upsilon_3$ are also LTI, and are given by the transfer functions $G_1(s) = \frac{2s+3}{(s+2)(s-\frac{2}{3})}$, $G_2(s) = \frac{2s-2.5}{(s+1)(s+2)}$, and $G_3(s) = \frac{2s-0.1}{(s-0.5)(s-0.4)}$, i.e. the first fault mode alters the poles of the nominal system, the second alters its zero, and the third alters both. It is easy to check that $\Upsilon_0$ is passive, that $\Upsilon_1$ is output $(-\frac{2}{3})$-passive, that $\Upsilon_2$ is input $(-1.25)$-passive, and that $\Upsilon_3$ is I/O $(-0.4,-0.4)$-passive, all with the storage function $S(x) = 0.5\|x\|^2$. Appropriately, we choose $\rho_0=\nu_0=0$; $\rho_1=-\frac{2}{3}$, $\nu_1=0$; $\rho_2 = 0$, $\nu_2 = -1.25$; $\rho_3=\nu_3 = -0.4$. We assume that all ten agents start with the nominal behaviour. Once every $5$ seconds, all agents independently change their behaviour, either to the nominal mode or to one of the fault modes. We assume that there is a chance of $p=0.75$ that an agent will follow the nominal mode in the next $5$ seconds, and a chance of $q=0.083$ that an agent will follow the $j$-the fault mode (for $j=1,2,3$). For edge controllers, we choose LTI controllers with transfer function $H(s) = \frac{1}{\tau s+1}$ for $\tau = 100$.
}

\textcolor{black}{
We first simulate the system without passivizing the agents, for a total of $150$ seconds. As the results in Fig. \ref{fig.Simulation_Nonpassivized} show, the agents do not synchronize, and in fact the distance of the agents' outputs from the mean can reach thousands near the end of the simulation. In order to remedy this problem, we look for a transformation $T$ which passivizes $\Upsilon_j$ for $j=0,1,2,3$. We find one by solving the following optimization problem:
\begin{align*}
\min_{T} \quad & (T_{11}-1)^2 + T_{12}^2 + T_{21}^2 + (T_{22}-1)^2 \\
{\rm s.t.}\quad & T S_{\rho_j,\nu_j} \ge 0,~ j=0,1,2,3
\end{align*}
where the matrix inequalities are understood component-wise. We choose this cost function as we wish to find a transformation $T$ which is as close as possible to the identity matrix (in the Frobenius norm), in order to have the smallest influence possible on the agents (while still passivizing them). This is a linearly-constrained least-squares problem which can be solved very quickly using off-the-shelf optimization software, e.g. Yalmip (\cite{Lofberg2004}). Solving the problem, we find the transformation $T \approx \left[\begin{smallmatrix} 0.6918 & 0.4622 \\ 0.4883 & 0.3896 \end{smallmatrix}\right]$, which takes about $1.3$ seconds. Applying the transformation for the modes $\Upsilon_j$, we get the following modes for the transformed agents.}

\textcolor{black}{
\vspace{-45pt}
\small
\begin{align*}
\tilde{\Upsilon}_0 &: \frac{0.7058(s+3.24)(s+1.356)}{(s+3.003)(s+1.334)},\\
\tilde{\Upsilon}_1 &: \frac{0.7058(s+2.506)(s+0.4232)}{(s+2.389)(s+0.2809)},
\end{align*}
\begin{align*}
\tilde{\Upsilon}_2 &: \frac{0.7058(s+4.595)(s+0.0011)}{(s+4.259)(s+0.0774)},\\
\tilde{\Upsilon}_3 &: \frac{0.7058(s+0.3774)(s+0.3185)}{(s+0.2181+0.2926\sqrt{-1})(s+0.2181-0.2926\sqrt{-1})},
\end{align*}\normalsize
all of which are (strictly) passive. We exemplify that the result of Proposition \ref{prop.FDINetwork} holds by simulating the network, where we take the same edge controllers as before, and implement the transformation $T$ as local controllers (as pre- and post-gains, constant gain feedback and constant gain feedthrough). The simulation length is $150$ seconds, as before. As seen in Fig. \ref{fig.Simulation_Passivized}, the output of the agents synchronizes at the $y=0$, as required.}

\section{Conclusions}
The paper considers the notion of $(\rho,\nu)$-passivity, which contains both shortage and excess of passivity. We characterized all I/O transformations mapping an I/O $(\rho,\nu)$-passive system to an I/O $(\rho_\star,\nu_\star)$-passive system. Starting with the SISO case, we used the geometric approach of \cite{Sharf2019a} to convert the problem into characterizing all linear transformations that map a given symmetric double-cone to a desired symmetric double-cone. We then solved the latter problem by studying the action of the collection of invertible $2\times 2$ matrices, $GL_2(\R)$, on the collection of symmetric double-cones. This culminated in a result showing that any I/O transformation mapping an I/O $(\rho,\nu)$-passive system to an I/O $(\rho_\star,\nu_\star)$-passive system can be written (up to a sign) as the product of three matrices, $S_{\rho,\nu}$, $S_{\rho_\star,\nu_\star}^{-1}$ and a non-negative matrix. We then shifted our focus to the MIMO case, where we showed a similar result, in which the non-negative matrix is replaced by a matrix satisfying a certain generalized version of an algebraic Riccati inequality. We presented three possible applications our results.
Future work can try and better characterize the collection of matrices satisfying the generalized algebraic Riccati inequality, giving a more explicit characterization for the MIMO case. Another avenue for future research can use the achieved parameterization to study various optimization problems, similar to the one discussed in Section \ref{sec.AppExam}.

\bibliographystyle{chicago}
\bibliography{main}             

\end{document}